\documentclass[a4paper,11pt]{article}
\usepackage{jheppub} 
\usepackage{lineno}
\usepackage[dvipsnames]{xcolor} 
\usepackage{dirtytalk}
\usepackage{ytableau}
\usepackage[dvipsnames]{xcolor}
\usepackage{dsfont}
\usepackage{mathdots}
\usepackage{colortbl}
\usepackage{braket}
\usepackage{amsthm}
\usepackage{amsmath}
\usepackage{float}
\usepackage{bbm, dsfont}
\usepackage[normalem]{ulem}
\usepackage{slashed}
\usepackage{mathtools}
\usepackage{multirow}
\usepackage{cancel}
\usepackage{cleveref}
\usepackage{orcidlink}
\usepackage{enumerate}
\usepackage{fontawesome}
\usepackage{adjustbox}
\usepackage{tcolorbox}
\usepackage{relsize}
\usepackage{csquotes}

\usepackage{tikz}
    \usetikzlibrary{positioning}
    \usetikzlibrary{arrows}
    \usetikzlibrary{shapes}
    \usepackage{pgfplots}
    \usetikzlibrary{calc}
    \usetikzlibrary{decorations.markings}
     \usetikzlibrary{decorations.pathmorphing}
    \tikzset{snake it/.style={decorate, decoration=snake}}

    \usetikzlibrary{matrix}

\pgfplotsset{compat=1.18}

\definecolor{ourblue}{RGB}{0, 57, 120} %
\colorlet{ourlightblue}{ourblue!50!white}
\newcommand{\bs}[1]{\boldsymbol{#1}}
\newcommand{\eps}{\varepsilon}
\newcommand{\rd}{\mathrm{d}}

\newcommand{\bx}{\bs{x}}

\newcommand{\bA}{\bs{A}}
\newcommand{\bZ}{\bs{Z}}
\newcommand{\bI}{\bs{I}}
\newcommand{\bJ}{\bs{J}}

\newcommand{\bS}{\bs{S}}

\newcommand{\bC}{\bs{C}}

\newcommand{\bM}{\bs{M}}

\newcommand{\bO}{\bs{O}}
\newcommand{\bR}{\bs{R}}
\newcommand{\bU}{\bs{U}}
\newcommand{\bW}{\bs{W}}

\newcommand{\bDelta}{\bs{\Delta}}
\newcommand{\bE}{\bs{E}}
\newcommand{\bSigma}{\bs{\Sigma}}
\newcommand{\bOmega}{\bs{\Omega}}

\newcommand{\zero}{{\scriptscriptstyle (0)}}
\newcommand{\one}{{\scriptscriptstyle (1)}}
\newcommand{\btau}{\bs{\tau}}

\def\beq{\begin{equation}}
\def\eeq{\end{equation}}
\def\bsp#1\esp{\begin{split}#1\end{split}}

\title{\boldmath Draft Template}

\author{Claude Duhr, Sara Maggio, Franziska Porkert, Cathrin Semper, Sven F. Stawinski}
\affiliation{Bethe Center for Theoretical Physics, Universit\"at Bonn, D-53115, Germany}

\preprint{BONN-TH-2025-24}

\emailAdd{cduhr@uni-bonn.de, smaggio@uni-bonn.de, fporkert@uni-bonn.de, csemper@uni-bonn.de,
sstawins@uni-bonn.de}

\title{Three-loop banana integrals with four unequal masses}
\abstract{We present a system of canonical differential equations satisfied by the three-loop banana integrals with four distinct non-zero masses in $D = 2-2\eps$ dimensions. Together with the initial condition in the small-mass limit, this provides all the ingredients to find analytic results for three-loop banana integrals in terms of iterated integrals to any desired order in the dimensional regulator. To obtain this result, we rely on recent advances in understanding the K3 geometry underlying these integrals and in how to construct rotations to an $\eps$-factorized basis. This rotation typically involves the introduction of objects defined as integrals of (derivatives of) K3 periods and rational functions. We apply and extend a method based on results from twisted cohomology to identify relations among these functions, which allows us to reduce their number considerably. We expect that the methods that we have applied here will prove useful to compute further multiloop multiscale Feynman integrals attached to non-trivial geometries.}

\begin{document}
\maketitle
\flushbottom


\section{Introduction}
\label{sec:introduction}

Feynman integrals are among the main quantities of interest in Quantum Field Theory (QFT). Not only are they the backbone of many, if not all, precise predictions for collider and gravitational wave experiments, but they are also an invaluable tool to study the mathematical structures underlying QFT. For this reason, the computation of multiloop Feynman integrals has received a lot of attention over the last decades. 

Despite their importance, the explicit computation of Feynman integrals is still a major bottleneck. One of the reasons is that Feynman integrals often diverge and need to be regulated. The most commonly used regularization scheme is dimensional regularization~\cite{tHooft:1972tcz}, where the integrals are computed in general dimension $D=d-2\eps$, and then expanded around some integer dimension $d$. The singularities of the integrals manifest themselves as poles in the dimensional regulator $\eps = \frac{d-D}{2}$.  

It has long been recognized that the Laurent coefficients are periods in the sense of Kontsevich and Zagier~\cite{MR1852188,Bogner:2007mn}. Thus, Feynman integrals are closely connected to geometry. The simplest class of periods that arise from Feynman integral computations are multiple polylogarithms (cf.,~e.g.,~refs.~\cite{Remiddi:1999ew,Goncharov:1998kja}). This class of special functions is by now well understood. An important step was the realization that dimensionally-regulated Feynman integrals satisfy so-called \emph{$\eps$-factorized} or \emph{canonical} differential equations, with only a linear dependence on the dimensional regularization parameter $\eps$~\cite{Henn:2013pwa}, allowing for a straightforward and algorithmic solution in terms of multiple polylogarithms.

Multiple polylogarithms are naturally connected to iterated integrals on the punctured Riemann sphere. It is well-known that multiple polylogarithms are insufficient to express all Feynman integrals starting from two-loop order, and functions associated to more complicated geometries also show up. The first geometry beyond the Riemann sphere appearing in Feynman integral computations is an elliptic curve and was identified in the context of the two-loop sunrise integral~\cite{Sabry,Caffo:1998du,Laporta:2004rb,Laporta:2008sx,Muller-Stach:2011qkg,Muller-Stach:2012tgj}. However, it took until 2015 for the relevant class of transcendental functions to be identified as elliptic polylogarithms~\cite{ell2,LevinRacinet,MR1265553,BrownLevin,Broedel:2014vla,ell15,EnriquezZerbini} and iterated integrals of modular forms~\cite{ManinModular,Brown:mmv,Adams:2017ejb}. Moreover, various proposals have been made how to extend the notions of {$\eps$-factorized} or {canonical} differential equations to dimensionally-regulated Feynman integrals associated to elliptic curves~\cite{Adams:2018yfj,Broedel:2018rwm,Bogner:2019lfa,Chen:2022lzr,Dlapa:2022wdu,Gorges:2023zgv,Chen:2025hzq,e-collaboration:2025frv,Chaubey:2025adn}. One of the novel features shared by all of these methods is that the differential equation matrix does not only involve algebraic functions like in the polylogarithmic case, but it also contains the periods of the elliptic curve, which are transcendental functions.

Elliptic curves do not exhaust the possible geometries that may arise from Feynman integrals, and we currently have examples of Feynman integrals attached to Calabi-Yau varieties~\cite{Brown:2010bw,Bloch:2014qca,Bloch:2016izu,Primo:2017ipr,Bourjaily:2018ycu,Bourjaily:2018yfy,Bourjaily:2019hmc,Broedel:2019kmn,Klemm:2019dbm,Bonisch:2020qmm,Bonisch:2021yfw,Duhr:2022dxb,Driesse:2024feo,Klemm:2024wtd,Frellesvig:2023bbf,Frellesvig:2024rea,Frellesvig:2024zph,Dlapa:2024cje,Forner:2024ojj,Duhr:2022pch,Duhr:2023eld,Duhr:2024hjf}, higher-genus curves~\cite{Huang:2013kh,Hauenstein:2014mda,Marzucca:2023gto,Abreu:2024jde} and special Fano geometries~\cite{Schimmrigk:2024xid}. Over the last couple of years, there has been a lot of activity in understanding how {$\eps$-factorized} or {canonical} differential equations can be extended to these kinds of geometries~\cite{Pogel:2022ken,Pogel:2022vat,Pogel:2022yat,Duhr:2024uid,Duhr:2025lbz,Maggio:2025jel}. A novel feature first observed in this context is the appearance of functions in the differential equation matrices that cannot only be expressed in terms of rational functions and periods of the underlying geometry, but also integrals over these functions are required. If and when these integrals can be evaluated in closed form in terms of periods and rational functions is still an open question.

The simplest and most-studied class of Feynman integrals associated to Calabi-Yau varieties are the so-called banana integrals. However, {$\eps$-factorized} or {canonical} differential equations have only been obtained for banana integrals where most propagator masses are equal~\cite{Pogel:2022ken,Pogel:2022vat,Pogel:2022yat,Duhr:2025lbz,Maggio:2025jel}. While various aspects of banana integrals with generic non-zero masses configurations have been studied~\cite{Bonisch:2020qmm,Bonisch:2021yfw,Duhr:2022dxb,Kreimer:2022fxm}, {$\eps$-factorized} or {canonical} differential equations have not yet been derived for these cases.\footnote{For alternative forms for the differential equations satisfied by banana integrals, see refs.~\cite{Mishnyakov:2023sly,Mishnyakov:2023wpd,Mishnyakov:2024rmb}.}

The main result of this paper is the construction of the canonical differential equations for the family of three-loop banana integrals with four distinct non-zero masses in $D=2-2\eps$ dimensions. In order to achieve this, we combine tools from various sources. Our main tool is the method of refs.~\cite{Gorges:2023zgv,Duhr:2025lbz,Maggio:2025jel} to construct a sequence of transformations that brings the system into $\eps$-factorized form. In this process, we draw heavily on the knowledge of the properties of the underlying Calabi-Yau geometry. For the three-loop banana integrals this is a family of K3 surfaces, and the corresponding geometry will be our second tool. Finally, we obtain an $\eps$-factorized system whose differential equation matrix involves 23 functions that are defined as (iterated) integrals over rational functions and (derivatives of) periods of the K3 surface. In order to simplify the differential equation matrix, we rely on results from twisted cohomology as a third tool. In particular, we exploit the fact that in a canonical basis the intersection matrix must be constant~\cite{Duhr:2024xsy}, which in turn imposes constraints on these new functions~\cite{Duhr:2024uid}. In this way we are able to express 10 of the 23 integrals in terms of periods and rational functions. Once symmetries related to permutations of the arguments are taken into account, we only require the definition of two functions that cannot be expressed in terms of rational and functions and K3 periods, thereby considerably simplifying the analytic expressions for the differential equation matrix. Together with the initial conditions (which we discuss in an appendix), this provides an analytic representation for all master integrals in terms of iterated integrals. Similar results have recently also been obtained in ref.~\cite{weinzierl_banana}.

This paper is organized as follows: In section~\ref{sec:setup} we introduce the three-loop banana integrals and the differential equations they satisfy. In section~\ref{sec:geometry} we review the geometry of the K3 surfaces attached to these integrals. In section~\ref{sec:canon} we show how to apply the method of refs.~\cite{Gorges:2023zgv,Duhr:2025lbz,Maggio:2025jel} to obtain canonical differential equations for the three-loop banana integrals with four distinct masses, and in section~\ref{sec:constrainingNewFcts} and~\ref{sec:analytic} we constrain the new functions that we had to introduce using methods from twisted cohomology and we present analytic solutions for them. In section~\ref{sec:conclusions} we draw our conclusions. We include various appendices where we discuss the initial conditions to the differential equations and where we collect expressions too lengthy to be shown in the main text. Mathematica files with the analytic expressions for the initial and the canonical differential equation matrices and the intersection matrix are available in computer readable form~\cite{bonndata}.


\section{The three-loop banana integral}
\label{sec:setup}
In section \ref{sec.def} we briefly define the family of integrals considered here 
and in section \ref{sec.defeq} we introduce its differential equation. 

\subsection{Definitions}
\label{sec.def}

\begin{figure}[!th]
\centering
\begin{tikzpicture}
\coordinate (llinks) at (-2.5,0);
\coordinate (rrechts) at (2.5,0);
\coordinate (links) at (-1.5,0);
\coordinate (rechts) at (1.5,0);
\begin{scope}[very thick,decoration={
    markings,
    mark=at position 0.5 with {\arrow{>}}}
    ] 
\draw [-, thick,postaction={decorate}] (links) to [bend right=25]  (rechts);
\draw [-, thick,postaction={decorate}] (links) to [bend right=-25]  (rechts);

\draw [-, thick,postaction={decorate}] (links) to [bend left=85]  (rechts);
\draw [-, thick,postaction={decorate}] (llinks) to [bend right=0]  (links);
\draw [-, thick,postaction={decorate}] (rechts) to [bend right=0]  (rrechts);
\end{scope}
\begin{scope}[very thick,decoration={
    markings,
    mark=at position 0.5 with {\arrow{>}}}
    ]
\draw [-, thick,postaction={decorate}] (links) to  [bend right=85] (rechts);
\end{scope}
\node (d1) at (0,1.1) [font=\scriptsize, text width=.2 cm]{$m_1$};
\node (d2) at (0,0.6) [font=\scriptsize, text width=.2 cm]{$m_2$};
\node (d3) at (0,-0.15) [font=\scriptsize, text width=.2 cm]{$m_3$};
\node (d4) at (0,-.7) [font=\scriptsize, text width=0.3 cm]{$m_4$};
\node (p1) at (-2.0,.25) [font=\scriptsize, text width=1 cm]{$p$};
\node (p2) at (2.4,.25) [font=\scriptsize, text width=1 cm]{$p$};
\end{tikzpicture}
\caption{The three-loop banana graph with four unequal masses.}
\label{fig:banana}
\end{figure}
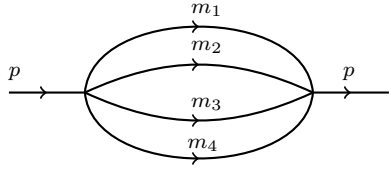

The main focus of our paper are three-loop banana integrals with four unequal masses,
which can be defined as the family of integrals (see fig.~\ref{fig:banana})
\begin{equation}
    I_{\nu_1,\ldots,\nu_9}=e^{3\gamma_E\eps}\int\left(\prod_{a=1}^{3}\frac{d^Dk_a}{i\pi^{\frac{D}{2}}}\right) \frac{1}{D_1^{\nu_1}\,D_2^{\nu_2}\,D_3^{\nu_3}\,D_4^{\nu_4}\,D_5^{\nu_5}\,D_6^{\nu_6}\,D_7^{\nu_7}\,D_8^{\nu_8}\,D_9^{\nu_9}}\,,
\end{equation}
with the propagators
\begin{align}
    &D_1=k_1^2-m_1^2\,,\qquad D_2=k_2^2-m_2^2\,,\qquad D_3=(k_1-k_3)^2-m_3^2\,,\nonumber\\
    & D_4=(k_2-k_3-p)^2-m_4^2\,,\qquad D_5=k_3^3\,,\qquad D_6=k_3\cdot p\,,\nonumber\\
    & D_7=k_1\cdot p\,,\qquad D_8=k_2\cdot p\,,\qquad D_9=k_1\cdot k_2\,.
\end{align}
We work in dimensional regularization~\cite{tHooft:1972tcz} in $D=2-2\eps$ dimensions, $\gamma_E=-\Gamma'(1)$ is the Euler-Mascheroni constant, and $\bs{\nu} = (\nu_1,\ldots,\nu_9)$ is a vector of integers. For the three-loop banana family we further impose $\nu_i\le 0$ for $i\ge 5$. Note that using dimensional-shift relations~\cite{Tarasov:1996br,Lee:2009dh}, we can relate these integrals to the banana integrals in $D=4-2\eps$ dimensions.

It is well known that the members of a family of Feynman integrals  are not all independent, but there are linear relations among them, called \emph{integration-by-parts} (IBP) relations~\cite{Tkachov:1981wb,Chetyrkin:1981qh}. Here they follow from the identity
\begin{equation}\label{eq:IBP_tot_diff}
\int \rd^D k_i\,\frac{\partial}{\partial k_i^\mu}\left(\frac{v^{\mu}}{D_1^{\nu_1}\dots D_9^{\nu_9}} \right)=0\,,
\end{equation}
where $v^\mu$ is an internal or external momentum. Since differentiation in the integrand produces propagators with shifted exponents, this identity leads to linear relations among different integrals from the same family, with coefficients that are rational functions of the kinematic variables and the dimensional regulator $\eps$. The IBP relations can be solved to express all integrals in the family in terms of a set of basis integrals, commonly referred to as \emph{master integrals} in the literature. The number of master integrals is always finite~\cite{Smirnov:2010hn,Bitoun:2017nre}. 

After solving the IBP identities using \texttt{LiteRed}~\cite{Lee:2012cn}, we identify 15 master integrals, which we collect into the vector ${\bs I} = (I_1,\ldots,I_{15})^T$. In the following we pick the basis:
\beq\bsp\label{bananbasis}
    I_1 &\, = I_{0,1,1,1,0,0,0,0,0}\,,\qquad  I_2  =     I_{1,0,1,1,0,0,0,0,0}\,,\qquad 
    I_3  =     I_{1,1,0,1,0,0,0,0,0}\,,\\
    I_4 &\, =     I_{1,1,1,0,0,0,0,0,0}\,,\qquad
    I_5  =     I_{1,1,1,1,0,0,0,0,0}\,,\qquad
        I_6  = 
    I_{2,1,1,1,0,0,0,0,0}\,,\\
    I_7 &\, =     I_{1,2,1,1,0,0,0,0,0}\,,\qquad
    I_8  =     I_{1,1,2,1,0,0,0,0,0}\,,\qquad
    I_9  =     I_{1,1,1,2,0,0,0,0,0}\,,\\
    I_{10} &\, =     I_{1,1,1,1,-1,0,0,0,0}\,,\qquad
        I_{11}  = 
    I_{1,1,1,1,0,-1,0,0,0}\,,\qquad
    I_{12}  =     I_{1,1,1,1,0,0,-1,0,0}\,,\\
    I_{13} &\, =     I_{1,1,1,1,0,0,0,-1,0}\,,\qquad
    I_{14}  =     I_{1,1,1,1,0,0,0,0,-1}\,,\qquad
    I_{15}  =     I_{3,1,1,1,0,0,0,0,0}
   \,.
\esp\eeq
There is of course some arbitrariness in how we choose a basis of master integrals. 
Our choice will be motivated in section~\ref{sec:canon}. Here we only note that there is a natural filtration on the vector space generated by the master integrals when the $\nu_i$ are strictly positive. Our choice of master integrals respects this filtration. In particular, $I_1,\ldots,I_4$ arise from subsectors and correspond to products of one-loop tadpole integrals. The remaining 11 integrals $I_5,\ldots,I_{15}$ are genuine banana integrals with $\nu_i>0$ for $i\le 4$. The integral $I_5= I_{1,1,1,1,0,0,0,0,0}$ will play a special role, and the integrals $I_6,\ldots,I_9$ and $I_{15}$ are derivatives of $I_5$,
\begin{align}
\label{dercorner}
    &\partial_{m_1^2}I_5=I_6\,,\qquad \partial_{m_2^2}I_5=I_7\,,\qquad \partial_{m_3^2}I_5=I_8\,,\qquad \partial_{m_4^2}I_5=I_9\,,\qquad 
    \partial^2_{m_1^2}I_5=2\,I_{15}\,.
\end{align}
The remaining master integrals $I_{10},\ldots,I_{14}$ are obtained by adding irreducible scalar products (ISPs) to the numerator of the integrand defining $I_5$.

\subsection{The system of differential equations satisfied by the three-loop banana integrals}
\label{sec.defeq}

To compute all members of a family of Feynman integrals, it is sufficient to compute its set of master integrals. One of the most commonly used ways to compute the master integrals is the method of differential equations~\cite{Kotikov:1990kg,Kotikov:1991hm,Kotikov:1991pm,Gehrmann:1999as,Henn:2013pwa}. For our integrals, the only non-trivial functional dependence of the integrals is through the variables $\bx=(x_1,x_2,x_3,x_4)$ defined as
\begin{equation} \label{defxi}
    x_i=\dfrac{m_i^2}{p^2}\,.
\end{equation}
For this reason, we may assume $p^2=1$ from now on. The vector of master integrals then fulfills a differential equation of the form 
\begin{equation}\label{eq:DEQ_prototype}
\rd \bI(\bx,\eps) = \bOmega(\bx,\eps)\bI(\bx,\eps)\,,
\end{equation}
where $\rd = \sum_{i=1}^4\rd x_i\,\partial_{x_i}$ denotes the exterior derivative with respect to the external kinematic parameters $\bx$.
The entries of the matrix $\bOmega(\bx,\eps)$ are rational functions of $\eps$ and rational one-forms in $\bx$, i.e., we have
\begin{equation}
\label{eq:startingDEQ}
\bOmega(\bx,\eps) = \sum_{i=1}^4\rd x_i\,\bOmega_i(\bx,\eps)= \sum_{i=1}^4\sum_{j=0}^3\rd x_i\,\eps^j\,\bOmega_i^{(j)}(\bx)\,,
\end{equation}
where the $\bOmega_i(\bx,\eps)$ are matrices of rational functions in $\bx$ and $\epsilon$. The explicit expressions for the matrices $\bOmega_i(\bx,\eps)$ are obtained by acting with the derivatives and using IBP relations. The result is lengthy and provided in computer
readable form \cite{bonndata}. Here we only mention that the differential equations respect the natural filtration given by the propagators. In particular, for our choice of basis in eq.~\eqref{bananbasis}, $\bOmega(\bx,\eps)$ takes the block-triangular form,
\beq
\bOmega(\bx,\eps) = \left(\begin{array}{ccc}
\bOmega_T(\bx,\eps) &&  0 \\
\bOmega_{I}(\bx,\eps) &\phantom{aa}& \bOmega_{\textrm{m.c.}}(\bx,\eps)
\end{array}\right)\,.
\eeq 
The matrix $\bOmega_T(\bx,\eps)$ is diagonal and describes the master integrals from the tadpole subsectors, and the matrix $\bOmega_{\textrm{m.c.}}(\bx,\eps)$ governs the differential equations for the maximal cuts of the banana integrals~\cite{Primo:2016ebd,Primo:2017ipr,Frellesvig:2017aai,Bosma:2017hrk}.

We have already mentioned that there is some level of arbitrariness in the choice of basis $\bI$.
This choice, however, may have an impact on our ability to solve the system of differential equations in eq.~\eqref{eq:DEQ_prototype}. If we make a different basis choice $\bI'$ related to the original basis through some rotation $\bM(\bx,\eps)$, i.e.,
\begin{equation}
    \bI'(\bx,\eps)=\bM(\bx,\eps) \bI(\bx,\eps) \,,
\end{equation}
then the differential equation matrix $\bOmega'(\bx,\eps)$ in the basis $\bI'(\bx,\eps)$ is related to the one in the original basis by a gauge transformation
\begin{equation}
    \bOmega'(\bx,\eps)=\Big(\bM(\bx,\eps)\bOmega(\bx,\eps)+\rd\bM(\bx,\eps)\Big)\bM(\bx,\eps)^{-1} \,.
\end{equation}
In the following we only consider rotations that respect the filtration defined by the propagators and that leave the master integrals from the subsectors invariant, i.e., we only consider rotations of the form
\beq
\bM(\bx,\eps) = \left(\begin{array}{cc}
\mathds{1} & {0}\\
* & *
\end{array}\right)\,.
\eeq
It is easy to check that $\bOmega'(\bx,\eps)$ is still block-triangular. Note that, even though $\bOmega(\bx,\eps)$ is a matrix of rational one-forms, there is no reason why $\bM(\bx,\eps)$, and thus $\bOmega'(\bx,\eps)$, must be rational. In fact, experience has shown that it is useful to consider non-rational matrices $\bM(\bx,\eps)$, because there is a distinguished set of bases
 in which the differential equation takes the form
\begin{equation}
    \rd\bJ(\bx,\eps)=\eps\bA(\bx)\bJ(\bx,\eps) \,,\qquad \bJ(\bx,\eps) = \bM(\bx,\eps)\bI(\bx,\eps) \,.
\end{equation}
Such bases are referred to as \emph{$\eps$-factorized} bases. There is an important subset of such bases, the so-called \emph{canonical} bases, which, in addition to being $\eps$-factorized, satisfy further nice properties (see below). These bases were first introduced in the context of Feynman integrals that evaluate to multiple polylogarithms~\cite{Remiddi:1999ew,Goncharov:1998kja}, in which case the matrix $\bA(\bx)$ is a matrix of $\rd\!\log$-forms. For Feynman integrals that evaluate to more complicated  classes of special functions, there is no clear consensus or definition of what a canonical basis is, apart from the fact that it must also be $\eps$-factorized (for an approach that advocates an alternative form to $\eps$-factorization, see ref.~\cite{Chaubey:2025adn}). Various different classes of bases have been proposed in the literature~\cite{Primo:2017ipr,Adams:2018yfj,Broedel:2018rwm,Bogner:2019lfa,Chen:2022lzr,Dlapa:2022wdu,Pogel:2022ken,Pogel:2022vat,Pogel:2022yat,Gorges:2023zgv,Chen:2025hzq,e-collaboration:2025frv} (for a comparison, see ref.~\cite{Frellesvig:2023iwr}). Here we rely on the operational definition of ref.~\cite{Gorges:2023zgv},\footnote{This method is known to deliver results that are equivalent to those based on the alternative approach introduced in refs.~\cite{Pogel:2022ken,Pogel:2022vat,Pogel:2022yat} (and presumably also ref.~\cite{e-collaboration:2025frv}), at least for the case of Feynman integrals associated to some classes of families of Calabi-Yau varieties~\cite{Duhr:2025lbz}.} which allows one to construct a rotation that achieves $\eps$-factorization. The method of ref.~\cite{Gorges:2023zgv} is inspired by a generalisation beyond polylogarithms~\cite{Broedel:2018qkq} of the concepts of \emph{uniform transcendental weight}~\cite{Kotikov:2010gf} and \emph{pure functions}~\cite{Arkani-Hamed:2010pyv}, which are the hallmark of a canonical basis in the polylogarithmic case. In addition, in all known cases the basis obtained from the construction of ref.~\cite{Gorges:2023zgv} enjoys additional nice properties which go beyond $\eps$-factorization. Based on these observations, it was proposed in ref.~\cite{Duhr:2025lbz} that a canonical basis should satisfy the following set of properties:
\begin{enumerate}
\item it is $\eps$-factorized,
\item the differential one-forms that span the matrix $\bA(\bx)$ are linearly independent in the sense that no constant linear combination is equal to a total derivative,
\item locally, the differential equation has at most logarithmic singularities.
\end{enumerate}
For a more in-depth discussion, in particular of the last two properties, we refer to ref.~\cite{Duhr:2025lbz}.

Three-loop banana integrals have been considered in the literature mostly in the case of equal propagator masses~\cite{Bloch:2014qca,Bloch:2016izu,Klemm:2019dbm,Bonisch:2020qmm,Bonisch:2021yfw,Broedel:2019kmn,Broedel:2021zij,Pogel:2022yat}. In ref.~\cite{Pogel:2022yat} a system of canonical differential equations in the above sense for the equal-mass three-loop banana integrals was obtained. Canonical bases were constructed in ref.~\cite{Maggio:2025jel} for the cases where either three of the four masses are equal, or the masses are equal pairwise (and for three equal masses, the maximal cuts are known to be expressible in terms of products of modular forms~\cite{Duhr:2025tdf,Duhr:2025ppd}). Banana integrals with four distinct non-zero masses, however, have not been studied at the same level. While series representations for (some of) the master integrals have been obtained~\cite{Bonisch:2020qmm,Bonisch:2021yfw}, there are currently no closed analytic expressions for all master integrals in terms of iterated integrals. In particular, no canonical basis has been constructed for this case. This can be traced back, on the one hand, to the fact that one has to deal with functions of four variables $\bx$, which tremendously increases the combinatorial and algebraic complexity, and, on the other hand, to the fact that the geometry underlying the three-loop banana integrals is more complicated.
 
 The main purpose of this paper is to provide a construction of a canonical basis for the complete set of master integrals of the three-loop banana integrals with four distinct non-zero masses in dimensional regularization. Before we do that in subsequent sections, we comment here more generally on the advantages of constructing a basis that satisfies the three properties outlined above.
 
 Let us start by discussing the consequence of the $\eps$-factorization. 
%
%
%
In an $\eps$-factorized basis the solution to the differential equation is readily given as
\begin{equation}
\bJ(\bx,\eps) = \bU_{\!\gamma}(\bx,\eps)\bJ_{\!0}(\eps)\,,
\end{equation}
where $\bJ_{\!0}(\eps)$ denotes the value of $\bJ(\bx,\eps)$ at some point $\bx=\bx_0$ and $\gamma$ is a path from $\bx_0$ to a generic point $\bx$. Here $\bs{U}_{\!\gamma}(\bs{x},\eps)$ denotes the path-ordered exponential 
\begin{equation}\label{eq:Pexp_def}
\bU_{\!\gamma}(\bx,\eps) = \mathbb{P}\exp\left[\eps\int_{\gamma}\bA(\bx)\right]\,.
\end{equation}
The path-ordered exponential can be expanded in $\eps$ up to any given order, and the coefficients of this expansion involve iterated integrals~\cite{ChenSymbol} built from the one-forms $\omega_i$, with $\bA(\bx) = \sum_{i=1}^p \bA_i \omega_i$, 
\begin{equation}\label{eq:Pexp_def_exp}
\bU_{\!\gamma}(\bx,\eps) = \mathds{1} + \sum_{k=1}^\infty \eps^k\sum_{1\le i_1,\ldots,i_k\le p}\bA_{i_1}\cdots\bA_{i_k}  I_{\gamma}(\omega_{i_1},\ldots,\omega_{i_k}) \,,
\end{equation}
where we defined the iterated integrals
\begin{equation}\label{eq:iterated_int_def}
\bsp I_{\gamma}(\omega_{i_1},\ldots,\omega_{i_k}) &\,= \int_{\gamma} \omega_{i_1}\cdots \omega_{i_k}\\
&\, = \int_{0\le \xi_k\le \cdots \le \xi_1\le1}\rd \xi_1 \,f_{i_1}(\xi_1)\,\rd \xi_2 \,f_{i_2}(\xi_2)\cdots \rd \xi_k\,f_{i_k}(\xi_k)\,,
\esp\end{equation}
where $\gamma^*\omega_{i_r} = \rd \xi_r\,f_{i_r}(\xi_r)$ is the pull-back of $\omega_{i_r}$ to the path $\gamma$. We see that the $\eps$-factorization translates into the fact that the coefficient of $\eps^k$ involves a linear combination with constant coefficients of iterated integrals of length $k$.

The remaining two properties have not been discussed as extensively in the literature. They make statements about the properties of the combinations of iterated integrals that arise from the path-ordered exponential. For example, the second property implies that the resulting iterated integrals are linearly independent~\cite{deneufchatel:hal-00558773,Duhr:2024xsy}, and there are no hidden zeroes in our combinations of iterated integrals. In section \ref{sec:constrainingNewFcts} we will see another consequence of this second defining property.

The third defining property of a canonical basis has implications for the choice of initial condition of the differential equation. In applications, natural choices for the initial condition $\bJ_{\!0}(\eps)$ are singular points $\bx_0$ of the differential equation. There are various methods to compute the asymptotic behaviour of Feynman integrals close to a singular point (cf., e.g., refs.~\cite{Smirnov:2002pj,Beneke:1997zp,Pak:2010pt,Jantzen:2011nz,Jantzen:2012mw,Ma:2023hrt,Jones:2024mfg,Chakraborty:2024uzz} and references therein). For the banana integrals a particularly convenient point is the small-mass limit $\bx\to0$, where explicit formulas can be derived~\cite{Bonisch:2021yfw,Iritani:2020qyh} (see also appendix~\ref{app:init_cond}). However, if $\bx=\bx_0$ is a singular point, then, strictly speaking, $\bJ(\bx,\eps)$ is ill-defined at $\bx=\bx_0$, and we need to use a regularization prescription to define $\bJ_{\!0}(\eps)$. This is most easily done in the case where $\bA(\bx)$ has at most logarithmic singularities at $\bx=\bx_0$, in which case we may interpret the singular point $\bx=\bx_0$ as a \emph{tangential base-point}~\cite{10.1007/978-1-4613-9649-9_3}, and there is a very natural way to define a regularized version of iterated integrals where the path $\gamma$ starts at this singular point $\bx_0$~\cite{Brown:mmv} (see also ref.~\cite{Abreu:2022mfk} for an exposition in a physics context).



\section{The geometry of three-loop banana integrals}
\label{sec:geometry}

The main goal of this paper is to present a differential equation in canonical form for the three-loop banana integrals with four distinct non-zero masses. Together with the initial conditions in the small-mass limit (see appendix~\ref{app:init_cond}), this furnishes an analytic representation for all master integrals in terms of iterated integrals. An important ingredient to construct the canonical basis is to understand the geometry attached to this class of integrals, which we review in this section.

It is by now well known that three-loop banana integrals are associated to a family of K3 surfaces~\cite{Bloch:2014qca,Bloch:2016izu,Klemm:2019dbm,Bonisch:2020qmm,Bonisch:2021yfw}. This is most apparent from the Baikov representation~\cite{BAIKOV1997347} of the maximal cuts of the integrals. Using a loop-by-loop approach~\cite{Frellesvig:2017aai,Frellesvig:2024ymq}, the maximal cut of the master integral $I_5$ can be cast in the form,
\begin{equation}\label{eq:Baikov}
    I_5^{\textrm{m.c.}}(\bx,\eps)=ie^{3\gamma_{\mathrm{E}}\eps}\left[\frac{\Gamma(1-\eps)}{\pi \Gamma(1-2\eps)}\right]^3\int_{\mathcal{C}}\rd z_1\rd z_2 \Psi(z_1,z_2;\bx) \,,
\end{equation}
where the precise form of the integration contour $\mathcal{C}$ is immaterial for our purposes, and we defined the multivalued function
\begin{align}
\nonumber\Psi&(z_1,z_2;\bx)= z_1^{\eps } (1+z_1+z_2)^{\eps } (4z_1-z_2^2)^{-\frac{1}{2}-\eps } \left[2 x_1 (x_3+z_1)-(x_3-z_1)^2-x_1^2\right]^{-\frac{1}{2}-\eps } \\
\label{banana4mtwist}&\times \left[2 x_2 (1+z_1+ z_2+x_4)-(1+z_1+z_2-x_4)^2-x_2^2\right]^{-\frac{1}{2}-\eps }\,,
\end{align}
where $z_1=k_3^2,\,z_2=2k_3\cdot p$. It can be shown that the algebraic variety defined by $y^2=\Psi(z_1,z_2;\bx)_{|\eps=0}$ defines a four-parameter family of K3 surfaces parametrized by the complex structure moduli $\bx$.\footnote{The family defined in this way is a slice of a more general family depending on more parameters. It is possible to identify another family, defined as complete intersections, which has the same periods and which depends precisely on four independent complex structure moduli. We refer to ref.~\cite{Bonisch:2020qmm} for a discussion.} In the following we give a brief review of the geometry of K3 surfaces, focusing on the aspects needed to understand the construction of the canonical basis in section~\ref{sec:canon}. For a detailed account of K3 surfaces, we refer to ref.~\cite{Huybrechts_2016}.

A K3 surface is a two-dimensional complex manifold $X$ with vanishing first Chern class. In the following we keep the discussion general, but we restrict to the case where $X$ depends on $d_X=4$ complex structure moduli $\bx$ (which is the case we are interested in). The Hodge diamond for a K3 manifold is 
\begin{align}\label{hodgediamond}
\begin{matrix}
    &&&1&&&\\
   &&0&&0&&\\
  1 &&& 20 &&&1\\
   &&0&&0&&\\
   &&&1&&&
\end{matrix}\,,
\end{align}
and its middle cohomology carries a pure Hodge structure of weight two,
\beq\label{eq:Hodge_decomp}
H^2(X,\mathbb{C}) = H^{2,0}(X)\oplus H^{1,1}(X)\oplus H^{0,2}(X)\,,
\eeq
where $H^{p,q}(X)$ contains cohomology classes generated by $(p,q)$-forms, i.e., differential forms containing exactly $p$ holomorphic and $q$ antiholomorphic differentials.
We then see that $X$ comes equipped with a distinguished holomorphic $(2,0)$-form $\Omega$.
We can also define the Hodge filtration on the middle cohomology,
\beq
F^2= H^{2,0}(X)\,,\qquad F^1=H^{2,0}(X)\oplus H^{1,1}(X)\,,\qquad F^0 = H^2(X,\mathbb{C})\,.
\eeq
Unlike the Hodge decomposition in eq.~\eqref{eq:Hodge_decomp}, the Hodge filtration varies holomorphically with the complex structure moduli $\bx$. In particular, Griffiths transversality states that
\beq\label{eq:Griffith}
\partial_{x_i}F^p\subseteq F^{p-1}\,.
\eeq

We can pair the holomorphic $(2,0)$-form $\Omega$ with cycles from the middle homology group $H_2(X,\mathbb{Z})$ to define the \emph{periods} of $X$. From eq.~\eqref{hodgediamond} we know that $H_2(X,\mathbb{Z})$ is 22-dimensional. One can show that $H_2(X,\mathbb{Z})$ contains a $20-d_X=16$-dimensional subspace of algebraic cycles, which give rise to vanishing periods. If we pick a basis $(\Gamma_0,\ldots,\Gamma_5)$ for the remaining six-dimensional space of transcendental cycles, we can define the vector of periods,
\begin{equation}\bsp
    \bs{\psi}(\bs{x})&\,=\big(\psi_0(\bs{x}),\psi_{1}^{(1)}(\bs{x}),\psi_{1}^{(2)}(\bs{x}),\psi_{1}^{(3)}(\bs{x}),\psi_{1}^{(4)}(\bs{x}),\psi_2(\bs{x})\big)^T\\
&\,=\Big(\int_{\Gamma_0}\Omega,\ldots,\int_{\Gamma_5}\Omega\Big)^T\,.
\esp\end{equation}

It is typically very hard to construct an explicit basis of transcendental cycles and to integrate $\Omega$ over them. Instead, the periods can be computed as solutions to a system of differential equations, which generate an ideal of differential operators called the \emph{Picard-Fuchs ideal} of $X$. An explicit set of generators for the Picard-Fuchs ideal for the three-loop banana integrals depending on four different masses was presented in ref.~\cite{Klemm:2019dbm,Bonisch:2020qmm}. 

The periods generally define multivalued functions of $\bx$. At the zero-mass point $\bx=0$, the periods of the K3 surface attached to the three-loop banana integral have a point of maximal unipotent monodromy (MUM), where the periods are given by 
\beq\bsp\label{eq:MUM-basis}
\psi_0(\bx) &\,= 1+\mathcal{O}(x_j)\,,\\
\psi_1^{(r)}(\bx) &\,= \psi_0(\bx)\,\log x_r+\mathcal{O}(x_j)\,,
\esp\eeq
and $\psi_2(\bx)$ is double-logarithmically divergent at the MUM-point (its precise form will be given below). 
In ref.~\cite{Bonisch:2020qmm}, a closed form for the expansion of $\psi_0$ close to $\bx=0$ was given.
\begin{align}
\psi_0(\bx) =&\sum_{n=0}^{\infty} \,\sum_{|k|=n} \binom{n}{k_1, \ldots, k_{4}}^2\, \prod_{i=1}^{4} x_i^{k_i} \,.
\end{align}

It is useful to introduce the following \emph{canonical coordinates} close to the MUM-point $\bx=0$,
\begin{equation}\label{deftaui}
\tau_r=\dfrac{\psi_1^{(r)}(\bx)}{\psi_0(\bx)} = \log x_r + \mathcal{O}(x_j)\,.
\end{equation}
Its inverse $\bx(\btau)$ is called the \emph{mirror map}. Later on we will need the derivatives of the mirror map, or equivalently the jacobian of the change of variables from $\btau=(\tau_1,\ldots,\tau_4)^T$ to $\bx$
\begin{equation}
\label{jacs}
    \bs{j}(\btau)=\Bigg(\frac{\partial x_i}{\partial{\tau_j}}\Bigg)_{1\le i,j\le 4}\,.
\end{equation}

The middle cohomology group of a K3 surface is equipped with a non-degenerate bilinear pairing, called the \emph{intersection pairing}. As a consequence of Griffiths transversality, the vector of periods satisfies the quadratic relations
\begin{equation}
\bs{\psi}^T(\bx)\bs{\Sigma}\,\bs{\psi}(\bx)=0\,\,,
\end{equation}
where $\bs{\Sigma}$ is the matrix that represents the intersection pairing in our basis. In particular, in the basis from eq.~\eqref{eq:MUM-basis}, the intersection matrix is given by
\begin{equation}\label{eq:quad_rel_K3}
\bs{\Sigma} = \begin{pmatrix}
 0 & 0& 1 \\
 0 & \bs{S} & 0 \\
 1 & 0& 0
 \end{pmatrix}\,, \textrm{~~~with~~~}
 \bs{S} = \begin{pmatrix}
  0 & -1 & -1 & -1  \\
  -1 & 0 & -1 & -1  \\
  -1 & -1 & 0 & -1  \\
  -1 & -1 & -1 & 0  
 \end{pmatrix}\,.
 \end{equation}
 We can use the quadratic relation in eq.~\eqref{eq:quad_rel_K3} to write $\psi_2(\bx)$ in terms of the other periods,
 \beq
 \psi_2(\bx) = -\frac{1}{2\psi_0(\bx)}\,\bs{\psi}^{(1)}(\bx)^T\bs{S}\bs{\psi}^{(1)}(\bx)\,,\qquad \bs{\psi}^{(1)}(\bx) =\big(\psi_{1}^{(1)}(\bs{x}),\ldots,\psi_{1}^{(4)}(\bs{x})\big)^T\,.
\eeq
The form of the intersection pairing in eq.~\eqref{eq:quad_rel_K3} allows one to identify the holomorphic period $\psi_0(\bx)$ as a hermitian modular form~\cite{Duhr:2025tdf}. 

It will be useful to consider the period matrix of the K3 surface, which can be chosen as
\beq\label{eq:W_matrix}
\bW(\bx) = \Big(\bs{\psi}(\bx), \partial_{x_1}\bs{\psi}(\bx),\ldots,\partial_{x_4}\bs{\psi}(\bx), \partial_{x_1}^2\bs{\psi}(\bx)\Big)^T\,.
\eeq
The period matrix also satisfies a set of quadratic relations
\beq\label{eq:quad_rels_W}
\bW(\bx)\bSigma\bW(\bx)^T = \bZ(\bx)\,,
\eeq
where $\bZ(\bx)$ is a matrix of rational functions (many entries are zero due to Griffiths transversality).


\section{A canonical basis for the three-loop banana integral}
\label{sec:canon}

In this section we detail our construction of the canonical basis for the three-loop banana integrals with four distinct non-zero masses in $D=2-2\eps$ dimensions. We use the method of ref.~\cite{Gorges:2023zgv} to construct a rotation to an $\eps$-factorized form. For all known examples~\cite{Duhr:2025lbz}, the resulting basis is also canonical in the sense defined in section~\ref{sec:setup}. 

An important starting point of the method of ref.~\cite{Gorges:2023zgv} is to identify a good initial basis of master integrals. It was already pointed out in ref.~\cite{Gorges:2023zgv} that this can often be achieved by performing an integrand analysis of the master integrals in integer dimensions. The subsectors corresponding to products of tadpole integrals are trivial, and we only discuss the top sector with $\nu_i>0$ for $i\le 4$. In ref.~\cite{Duhr:2025lbz} it was argued that the initial basis should be chosen such that it aligns with the mixed-Hodge structure (MHS)~\cite{PMIHES_1971__40__5_0,PMIHES_1974__44__5_0} of the geometry associated to the maximal cuts at $\eps=0$. As reviewed in the previous section, in our case the relevant geometry is a four-parameter family of K3 surfaces, whose Hodge structure is given in eq.~\eqref{eq:Hodge_decomp}. The entries of the period vector $\bs{\psi}(\bx)$ compute the maximal cuts of the master integral $I_5$ at $\eps=0$. Following the argument from ref.~\cite{Duhr:2025lbz}, we then see from eq.~\eqref{dercorner} and from Griffiths transversality in eq.~\eqref{eq:Griffith} that the maximal cuts of $I_6,\ldots,I_9$ and $I_{15}$ can be related to the derivatives of $\bs{\psi}(\bx)$. Hence, we see that our initial choice of master integrals $I_5,\ldots,I_9$ and $I_{15}$ is aligned with the pure Hodge structure of the associated K3 surface. The remaining master integrals $I_{10},\ldots,I_{14}$ are obtained by adding ISPs to the numerators of the integrand of $I_5$. We therefore expect that they have non-vanishing singularities at infinity, and they should be associated to higher-weight contributions in the MHS. This motivates a posteriori our choice of basis of master integrals from eq.~\eqref{bananbasis}.

After identification of an appropriate initial basis aligned with the MHS, the procedure of ref.~\cite{Gorges:2023zgv} is to construct a sequence of rotations to achieve $\eps$-factorization. More precisely, the canonical basis of master integrals $\bJ(\bx,\eps)$ is related to the initial basis $\bI(\bx,\eps)$ from eq.~\eqref{bananbasis} through a sequence of transformations
\beq\label{eq:J_can}
\bJ(\bx,\eps) = \bU_{\!\textrm{t}}(\bx,\eps)\bU_{\!{\eps}}(\eps)\bU_{\!\textrm{ss}}(\bx)\bI(\bx,\eps)\,,
\eeq
where $\bU_{\!\textrm{t}}(\bx,\eps)$, $\bU_{\!{\eps}}(\eps)$ and $\bU_{\!\textrm{ss}}(\bx)$ are $15\times15$ matrices.
We now sketch the construction of each transformation in turn, focusing on the aspects specific to the case of the three-loop banana integrals. For a general discussion of the different steps, we refer to refs.~\cite{Gorges:2023zgv,Duhr:2025lbz}.

The first transformation $\bU_{\!\textrm{ss}}(\bx)$ is constructed as follows: Consider the period matrix $\bW(\bx)$ of the K3 surface for $\eps=0$ defined in eq.~\eqref{eq:W_matrix}. It may be split into a semi-simple and unipotent part~\cite{Broedel:2018qkq},
\beq
\bW(\bx) = \bW^{\textrm{ss}}(\bx)\bW^{\textrm{u}}(\bx)\,.
\eeq
For a procedure how this splitting can be achieved for K3 surfaces, we refer to refs.~\cite{Broedel:2019kmn,Ducker:2025wfl,Duhr:2025lbz,Maggio:2025jel}. We note that the quadratic relations among periods and their derivatives in eq.~\eqref{eq:quad_rels_W} play an important role here. In particular, in ref.~\cite{Duhr:2025lbz} it was shown that the quadratic relations in eq.~\eqref{eq:quad_rels_W} can be restricted to the semi-simple part,\footnote{In ref.~\cite{Duhr:2025lbz} this was strictly  shown only for one-parameter families of K3 surfaces. It is easy to check that the proof extends to our case.}
\begin{equation}\label{CYZ}
   \bW^{\textrm{ss}}(\bx) \bs{\Sigma}\bW^{\textrm{ss}}(\bx)^T = \bZ(\bx)\,.
\end{equation}
 For the subsectors and the integrals $I_{10},\ldots, I_{14}$, this procedure is trivial (if the integrals are normalised to have unit leading singularities~\cite{Cachazo:2008vp,Arkani-Hamed:2010pyv}). In the end, we define the rotation matrix $\bU_{\!\textrm{ss}}(\bx)$ in eq.~\eqref{eq:J_can} by
\begin{equation}
\bU_{\!\textrm{ss}}(\bx)=\left(
    \begin{array}{ccccccccccccccc}
      \mathds{1}_{4\times 4}&0&0 &0&0\\
       0&\frac{1}{\psi_0}&0&0&0\\
       0&\partial_{\bs{\tau}}\frac{1}{\psi_0}&\frac{\bs{j}^T}{\psi_0} &0&0\\
0&0 &0&\mathds{1}_{5\times 5}&0\\
0&f_1&\bs{f}&0&f_6
    \end{array}\right)\,,
\end{equation}
where the jacobian $\bs{j}$ was defined in eq.~\eqref{jacs} and $\bs{f}=(f_2,f_3,f_4,f_5)$. The functions $f_1,\ldots,f_6$ are determined from the quadratic relations for the semi-simple part in eq.~\eqref{CYZ}.

The second transformation~$\bU_{\!{\eps}}(\eps)$ is independent of $\bx$ and only adjusts the $\eps$ scaling. In our case, it is given by
\begin{equation}
\bU_{\eps}(\eps)=\left(
    \begin{array}{ccccccccccccccc}
        \eps^3 \mathds{1}_{5\times 5} & 0 & 0 & 0  \\
        0 & \eps^2 \mathds{1}_{4\times 4} & 0 & 0 \\
        0 & 0&\eps^3 \mathds{1}_{5\times 5} & 0 \\
        0 & 0 & 0 & \eps \\
    \end{array}\right)\,.
\end{equation}
The choice for this particular scaling takes inspiration from the generalization of the concept of uniform transcendental weight beyond polylogarithms from ref.~\cite{Broedel:2018qkq}.

After having performed the first two transformations, the differential equation matrix is in a form where all entries that are not yet $\eps$-factorized are below the diagonal. We can then achieve full $\eps$-factorization through a final transformation $\bU_{\!\textrm{t}}(\bx,\eps)$.
In our case, we find that this matrix takes the form
\beq\label{Ut4massbanana}
\bU_{\!\textrm{t}}(\bx,\eps) = \frac{1}{\eps}\bU_{\!\textrm{t}}^{(-1)}(\bx) + \bU_{\!\textrm{t}}^{(0)}(\bx)\,,
\eeq
with
\beq\label{eq:Utm1}
\bU_{\!\textrm{t}}^{(-1)}(\bx) = \begin{pmatrix}
\mathds{1}_{4\times 4} & {0} & {0} & {0} \\ 
 {0} & 1&  {0} & {0} \\ 
 {0} & {0} &\mathds{1}_{9\times 9} & {0}  \\ 
 {0} &G_0(\bx) & {0} & 1
 \end{pmatrix}\,.
 \eeq
For later convenience, we parametrize the matrix $\bU_{\!\textrm{t}}^{(0)}(\bx)$ as
\beq\label{eq:Ut0}
 \bU_{\!\textrm{t}}^\zero(\bx) = \begin{pmatrix}
 \mathds{1}_{4\times 4} & 0\\
 \bU_{\textrm{st}}^\zero(\bx)& \bU_{\textrm{m.c.}}^\zero(\bx)\end{pmatrix}\,,
\eeq
with
\beq\bsp\label{eq:U0mc}
 \bU_{\textrm{st}}^\zero(\bx) &\,= \begin{psmallmatrix}
 0&0&0&0\\
 \vdots&\vdots&\vdots&\vdots\\
 0&0&0&0\\
 t_{15,1}(\bx)&  t_{15,2}(\bx) &  t_{15,3}(\bx)&  t_{15,4}(\bx)
 \end{psmallmatrix}\textrm{~~~and~~~}
 \bU_{\!\textrm{m.c.}}^\zero(\bx)=
\begin{pmatrix}
1& \mathbf{0} &0\\
\bs{t}_{5}(\bx)^T&\mathds{1}_{9\times9} & \bs{0} \\
t_{15,5}(\bx)&\mathbf{t}_{15}(\bx)& 1
\end{pmatrix}\, 
\esp\end{equation}
and we defined
\beq\bsp\label{eq:tvec_def}
\mathbf{t}_{5}(\bx) &\,=
\begin{pmatrix}
t_{6,5}(\bx) & \cdots & t_{14,5}(\bx)
\end{pmatrix}\,,\\
\mathbf{t}_{15}(\bx) &\,=
\begin{pmatrix}
t_{15,6}(\bx) & \cdots & t_{15,14}(\bx)
\end{pmatrix}\,.
\esp\eeq
The functions $t_{i,j}(\bx)$ and $G_0(\bx)$ in the non-trivial entries of eqs.~(\ref{eq:Utm1} - \ref{eq:tvec_def}) are not fixed a priori, but they are determined in such a way that the vector $\bJ(\bx,\eps)$ from eq.~\eqref{eq:J_can} satisfies a differential equation in $\eps$-factorized form. This leads to a set of first-order differential equations for the $t_{i,j}(\bx)$ and $G_0(\bx)$. Note that only the functions $t_{i,j}(\bx)$ enter the differential equation matrix $\bA(\bx)$, and the function $G_0(\bx)$ can be expressed as derivatives involving the $t_{i,j}(\bx)$~\cite{Duhr:2025lbz} (see also section~\ref{solnewfunctions}).

In the following we refer to the functions $t_{i,j}(\bx)$ and $G_0$ as \emph{$\eps$-functions}. We can solve the differential equations defining the $\eps$-functions in two ways. First, we have determined series expansions from the differential equation for all the $t_{i,j}(\bx)$ and $G_0$ close to the MUM-point (see appendix~\ref{app:series}). This allows us to evaluate these functions (at least locally close to $\bx=0$), thereby providing an explicit representation of the canonical differential equation matrix. Second, the solutions to the differential equations can be cast in the form of iterated integrals over kernels that involve rational functions as well as the K3 periods and their derivatives. However, unlike the iterated integrals that may arise as solutions to the canonical differential equations, these iterated integrals are \emph{not} multivalued at the MUM-point $\bx=0$, but instead they admit Laurent expansions close to the MUM-point.  We will discuss the analytic expressions of the $\eps$-functions in terms of iterated integrals in section~\ref{sec:analytic}.

We have checked that the matrix $\bA(\bx)$ only has logarithmic singularities at the MUM-point $\bx=0$. We stress that this was not used as an input to the construction of the transformation $\bU(\bx,\eps)$, corroborating the expectation of ref.~\cite{Duhr:2025lbz} that the construction of ref.~\cite{Gorges:2023zgv} (and also those of refs.~\cite{Pogel:2022ken,Pogel:2022vat,Pogel:2022yat,Maggio:2025jel,e-collaboration:2025frv}, which are expected to deliver equivalent bases) goes beyond mere $\eps$-factorization and produces bases that can be called `canonical' according to the definition of section~\ref{sec:setup}. We will comment on the third defining property in section~\ref{sec:constrainingNewFcts}.

As explained in section~\ref{sec:setup}, the fact that $\bA(\bx)$ only has logarithmic singularities also has practical implications. In appendix~\ref{app:init_cond} we show how to determine the initial condition of the differential equation at the zero-mass point $\bx=0$. This is a singular point of the differential equation (in fact, it is a MUM-point), but we can use a tangential base-point regularisation to define iterated integrals where the path $\gamma$ starts at the singular point $\bx_0=0$. In this way we can obtain analytic results for all master integrals in terms of iterated integrals over the integration kernels that define the matrix $\bA(\bx)$. Since we have series representations for the functions $t_{i,j}(\bx)$, we can evaluate these iterated integrals close to the MUM-point $\bx=0$. We stress that numerical results and series representations close to the MUM-point for some of the master integrals have already been described in ref.~\cite{Bonisch:2021yfw}.

The differential equation matrix $\bA(\bx)$, and thus also the kernels that define the iterated integrals, involve the 23 $\eps$-functions $t_{i,j}(\bx)$. Before we discuss analytic representations of these in terms of iterated integrals, it is an interesting question to ask if there are relations between these integrals and/or if a subset of them can be evaluated in terms of other functions, for example rational functions and (derivatives of) periods of the K3 surface. This question will be addressed in the next section.


\section{Constraining the \texorpdfstring{$\eps$}{eps}-functions}
\label{sec:constrainingNewFcts}

In this section we show how we can obtain closed expressions for 10 of the 23 $\eps$-functions $t_{i,j}(\bx)$ introduced in the previous section in terms of rational functions and (derivatives of) periods of the underlying K3 surface.
Our strategy closely follows the ideas introduced in refs.~\cite{Duhr:2024uid,Duhr:2024xsy,paper1}, in particular the observation that the twisted cohomology intersection matrix in a canonical basis is expected to be constant. In addition, maximal cuts exhibit a certain natural notion of self-duality~\cite{Duhr:2024xsy}, which can then be used to constrain the matrix of differential equations for the maximal cut~\cite{Duhr:2024uid,paper1}. For a closely related idea, see also ref.~\cite{Pogel:2024sdi}.

\subsection{Twisted cohomology and self-duality}
\label{sec:twisted_review}

Twisted cohomology is a framework to study integrals of the form
\begin{equation}
\label{eq:twistedIntegral}
    \int_{\mathcal{C}}\Psi\,\varphi \,,
\end{equation}
where $\Psi$ is a multivalued function, the so-called \emph{twist}, $\varphi$ is a differential form and $\mathcal{C}$ is a (twisted) cycle, which will play no role in our considerations. From the Baikov representation for example (cf.~eq.~\eqref{eq:Baikov}), it is clear that Feynman integrals in dimensional regularization match this form, and so they can be studied in the twisted cohomology framework~\cite{Mastrolia:2018uzb}. In the following, we will briefly introduce some necessary concepts from the theory of twisted cohomology. For a more complete treatment and further mathematical details we refer to ref.~\cite{aomoto_theory_2011}.

For a fixed twist, the set of differential forms $\varphi$ leading to well-defined integrals as in eq.~\eqref{eq:twistedIntegral} form a finite-dimensional vector space, the so-called \emph{twisted cohomology group}. If we pick a basis of this vector space, then the integrals in eq.~\eqref{eq:twistedIntegral} correspond, roughly speaking, to the master integrals of the Feynman integral family.\footnote{There are some differences, for example due to relations between integrals which do not hold on the level of integrands. In the case of the unequal-mass banana integral studied in this paper this distinction is not necessary and we will hence not discuss it further, see however, e.g.,~refs.~\cite{Chen:2022lzr,Gasparotto:2023roh,e-collaboration:2025frv,paper3}.}  In particular the basis forms of the twisted cohomology group satisfy the same differential equation \eqref{eq:DEQ_prototype} as the master integrals.

Furthermore, there is a vector space of differential forms $\varphi$ corresponding to the twist $\Psi^{-1}$, called the the \emph{dual} twisted cohomology group. For maximally cut Feynman integrals, given a basis of master integrals (i.e., a basis for the twisted cohomology group), one may choose a basis of dual maximal cut Feynman integrals~\cite{Caron-Huot:2021xqj,Caron-Huot:2021iev} (i.e., a basis for the dual twisted cohomology group) such that the integrals and their duals only differ by replacing $\eps$ with $-\eps$~\cite{Duhr:2024rxe}. We refer to this as \emph{self-duality}.

There is a bilinear pairing between the twisted cohomology group and its dual. The Gram matrix arising from this pairing is called the (twisted cohomology) \emph{intersection matrix} $\bC(\bx,\eps)$. For details on the definition of the pairing and how the intersection matrix can be computed, see refs.~\cite{Frellesvig:2019kgj,Frellesvig:2019uqt,Weinzierl:2020xyy,Frellesvig:2020qot,Chestnov:2022xsy,Fontana:2023amt,Brunello:2023rpq,Brunello:2024tqf} and references therein. Here let us just remark that the intersection matrix $\bC(\bx,\eps)$ is a matrix of rational functions in the kinematical parameters and the dimensional regulator, and that it can be computed (at least in principle) in an algorithmic way.
In the presence of self-duality, it satisfies a differential equation
\begin{equation}\label{deqC}
\rd \bC(\bx,\eps)=\bOmega(\bx,\eps) \bC(\bx,\eps)+\bC(\bx,\eps) \bOmega(\bx,-\eps)^T\,.
\end{equation}


\subsection{Constraining \texorpdfstring{$\eps$}{eps}-functions in the top sector}
\label{sec:constraints}

We now discuss how ideas from twisted cohomology can be applied to express some of the $\eps$-functions in terms of rational functions, K3 periods, and their derivatives. The method will rely on the self-duality (in the sense of twisted cohomology) of the maximal cuts~\cite{Duhr:2024rxe}. For this reason, our approach is limited to the top sector describing the maximal cuts of the banana integrals. To make this explicit, we parametrize the canonical differential equation matrix $\bA(\bx)$ in the form:
\beq
\bA(\bx) = \begin{pmatrix}
\bs{D}_{\textrm{tad.}}(\bx)& 0\\
 \bs{B}(\bx)& \bs{A}_{\textrm{m.c.}}(\bx)\end{pmatrix}
 \,.
\eeq
Note that this decomposition matches the decomposition of $\bU_{\!\textrm{t}}^\zero(\bx)$ in eq.~\eqref{eq:Ut0}.
%
%
%
We also introduce the matrices $\bU_{\!\textrm{m.c.}}(\bx,\eps)$ and $\bU_{\!\textrm{st}}(\bx,\eps)$ through a similar decomposition for $\bU(\bx,\eps)$.
In the remainder of this section we will only discuss the top sector, i.e., we will only focus on the vector of maximal cuts 
\beq
\bI_{\textrm{m.c.}}(\bx,\eps) = \Big(I_5^{\textrm{m.c.}}(\bx,\eps),\ldots,I_{15}^{\textrm{m.c.}}(\bx,\eps)\Big)^T\,.
\eeq
 We will also define the vector of maximal cuts in the canonical basis
\beq
\bJ_{\textrm{m.c.}}(\bx,\eps) =  \bU_{\!\textrm{m.c.}}(\bx,\eps)\bI_{\textrm{m.c.}}(\bx,\eps)\,.
\eeq
It satisfies the differential equation
\beq
\rd\bJ_{\textrm{m.c.}}(\bx,\eps) =\eps\bA_{\textrm{m.c.}}(\bx,\eps)\bJ_{\textrm{m.c.}}(\bx,\eps) \,.
\eeq

We now interpret our basis of master integrals as a basis for the twisted cohomology group attached to the three-loop banana integrals. Since we focus on maximal cuts, we are in a self-dual situation.
We start from the observation of ref.~\cite{Duhr:2024xsy} that, if we choose a basis such that the differential equation is $\eps$-factorized and the differential one-forms in the matrix $\bA(\bx)$ are linearly independent (in the sense defined in section~\ref{sec:setup}), then the intersection matrix between the differential forms in that basis is always constant. Note that these two conditions are satisfied for canonical bases, and so the intersection matrix in a canonical basis is constant. We have computed the intersection matrix $\bC(\bx,\eps)$ for the maximal cuts of the three-loop banana integrals in the initial basis $\bI_{\textrm{m.c.}}(\bx,\eps)$ using the method of ref.~\cite{Frellesvig:2020qot} in the loop-by-loop Baikov representation with the twist in eq.~\eqref{banana4mtwist}. The expressions for the differential forms corresponding the the ISPs integrated out in the loop-by-loop Baikov approach can be obtained as described in ref.~\cite{Frellesvig:2019kgj,Frellesvig:2024ymq}. The algorithm of ref.~\cite{Frellesvig:2020qot} relies on solving systems of intermediate linear equations. Although the final expression may be comparably simple, the intermediate complexity for some of the intersection numbers is significant and the simplification of the resulting sums requires considerable computational effort or additional techniques to control the growth of the expressions. To mitigate the complexity, we only compute a subset of the intersection numbers using the algorithm of  ref.~\cite{Frellesvig:2020qot}--especially those entries we expect to vanish or to be constant. We then make an ansatz for the intersection matrix keeping the undetermined entries general. Then, we made use of the differential equation~\eqref{deqC} in the following fashion: Assume that we have determined using the algorithm of ref.~\cite{Frellesvig:2020qot} that the element $\bC(\bx,\eps)_{ij}$ is constant in $\bx$. Then we obtain
\begin{equation}
0=\rd \bC(\bx,\eps)_{ij}=\Big[\bOmega(\bx,\eps) \bC(\bx,\eps)+\bC(\bx,\eps) \bOmega(\bx,-\eps)^T\Big]_{ij}\,.
\end{equation}
We interpret this equation as a linear system relating some of the entries of the intersection matrix. In our case, this was enough to determine all the remaining entries of $\bC(\bx,\eps)$. Note that $\bC(\bx,\eps=0)$ corresponds to $\bs{Z}(\bx)$ introduced in eq.~\eqref{eq:quad_rels_W} if we delete the rows and columns that are associated to the ISPs. 
The intersection matrix $\bC(\bx,\eps)$ is an $11\times11$ matrix of rational functions in $\bx$ and $\eps$. The expressions are lengthy and provided in computer readable form~\cite{bonndata}. 

If we change basis to the canonical basis $\bJ_{\textrm{m.c.}}(\bx,\eps)$ for the maximal cuts, the intersection matrix becomes,
\begin{equation}\label{eq:constant_intersection}
    \bU_{\!\textrm{m.c.}}(\bx,\eps)\bC(\bx,\eps)\bU_{\!\textrm{m.c.}}(\bx,-\eps)^T=-\frac{\eps ^4}{4}\,\bDelta_{} \,,
\end{equation}
with 
\beq
\bDelta= \begin{pmatrix}
    0 & \phantom{-}0 & \phantom{-} 0 &\phantom{-}1 \\
    0 & -\bS &\phantom{-} 0 &\phantom{-}0 \\
    0 & \phantom{-}0 &\phantom{-} \bE &\phantom{-}0 \\
    1 & \phantom{-}0 &\phantom{-} 0 &\phantom{-}0 \\
\end{pmatrix}\,,
\label{delta4masses}
\eeq
where $\bS$ is defined in eq.~\eqref{eq:quad_rel_K3} and 
\begin{equation}
\bE =\frac{1}{24} 
\begin{pmatrix}
\phantom{-}12  & -4  & -2& -2 &  \phantom{-}2 \\
-4 & \phantom{-}4  & \phantom{-}2 & \phantom{-}2 & \phantom{-}0 \\
-2 & \phantom{-}2 &  \phantom{-}3  &  \phantom{-}1 & \phantom{-}1 \\
-2 & \phantom{-}2 &  \phantom{-}1 & \phantom{-}3  & -1 \\
\phantom{-}2 & \phantom{-}0 &  \phantom{-}1 & -1 &  \phantom{-}3\\
\end{pmatrix}.
\end{equation}
We see that, as expected, the intersection matrix is $\bx$-independent in the canonical basis. Let us make two comments at this point. First, we note that the fact that the $\eps$-dependence factorizes in the right-hand side of eq.~\eqref{eq:constant_intersection} is not a coincidence, but this is a general feature for maximal cuts (and follows more generally from self-duality)~\cite{Duhr:2024xsy}. Second, we stress that the result of ref.~\cite{Duhr:2024xsy} requires not only $\eps$-factorization, but also the linear independence of the differential forms. The method of ref.~\cite{Gorges:2023zgv}, however, only guarantees the construction of a matrix $\bU_{\!\textrm{m.c.}}(\bx,\eps)$ that achieves  $\eps$-factorization. As already mentioned, in all cases the resulting basis also enjoys linear independence of the differential forms~\cite{Duhr:2025lbz}. The fact that we find a constant intersection matrix confirms this expectation yet again.

In ref.~\cite{Duhr:2024xsy} it was shown that the constancy of the intersection matrix in a canonical basis implies a certain symmetry of the differential equation matrix $\bA_{\textrm{m.c.}}(\bx)$,
\beq\label{eq:DEQ_symmetry}
\varphi_{\Delta}\Big(\bA_{\textrm{m.c.}}(\bx)\Big) = \bA_{\textrm{m.c.}}(\bx)\,,
\eeq
where the map $\varphi_{\Delta}$ is called the \emph{$\Delta$-transposition} and acts on matrices $\bM$ via
\beq
\varphi_{\Delta}(\bM) = \bDelta\bM^T\bDelta^{-1}\,.
\eeq
Note that $\bA_{\textrm{m.c.}}(\bx)$ depends on the $\eps$-functions $t_{i,j}(\bx)$ in eq.~\eqref{eq:U0mc}.
In ref.~\cite{Duhr:2024uid} the symmetry in eq.~\eqref{eq:DEQ_symmetry} was used to derive relations between the $\eps$-functions $t_{i,j}(\bx)$. In our case, some examples of these relations read
\begin{align}
\label{4massesrel1}
\sum_{k=1}^4 \frac{\bC^\one_{k+1,11}\, \bs{j}_{k,i-5}}{\bC_{1,11} }&= \,t_{i,5}+\sum_{j=6,\,j\neq i}^9 \,t_{15,j}\,,\quad i\in \{6,7,8,9\} \,,\\
\label{4massesrel4}
\frac{ \psi_0^2 \bC^{\scriptscriptstyle(2)}_{11,11}}{8(\bC_{1,11})^2}&=\,t_{15,5}+ \sum_{i=6}^{13}\,t_{i,5} \,t_{15,i}+t_{15,6}\sum_{i=7}^9 t_{15,i}+\frac{t_{15,11}}{12}\sum_{i=11}^{13}t_{15,i}\\
&+\,t_{15,7} \,t_{15,8}+\left(\,t_{15,7}+\,t_{15,8}\right) \,t_{15,9}-\frac{1}{16}\left(\,t_{15,12}^2- \,t_{15,13}^2
- \,t_{15,14}^2\right)\notag\\
&-\frac{1}{12} \,t_{15,10} \left(2 \,t_{15,11}+\,t_{15,12}+\,t_{15,13}-\,t_{15,14}\right)-\frac{1}{4} \,t_{15,10}^2\notag\\
&
+\frac{1}{24} \left(24 \,t_{14,5}+\,t_{15,12}-\,t_{15,13}\right) \,t_{15,14}+\frac{1}{24} \,t_{15,12} \,t_{15,13}
\Big)\notag\,.
\end{align}
Here $\bC^{(i)}$ refers to the coefficient of $\eps^i$ in the Laurent expansion of  $\bC$,
\beq
\bC(\bx,\eps) = \sum_{i\ge i_0}\bC^{(i)}(\bx)\eps^i\,.
\eeq
The complete set of relations is provided in computer readable form~\cite{bonndata}. We see that some of the relations are linear, and thus easy to solve. Others, however, are non-linear. This does not make it straightforward to find a good parametrization of the $\eps$-functions that can be fixed in terms of rational functions and periods. We will study this issue in detail in ref.~\cite{paper1}, where also rigorous mathematical proofs will be presented that hold for general $\bDelta$. Here we content ourselves to present the results that are specific to the case of the three-loop banana integrals.

As a starting point, consider the group $G$ of all unipotent lower-triangular matrices that is closed under $\Delta$-transposition for our particular choice of $\bDelta$. It is easy to check that the matrix $\bU_{\textrm{m.c.}}^\zero(\bx)$ in eq.~\eqref{eq:U0mc} defines an element in $G$.
In ref.~\cite{paper1} we will show very generally that every element $\bM\in G$ admits a unique decomposition of the form
\beq\label{eq:OR}
\bM = \bO\bR\,,
\eeq
where $\bO$ and $\bR$ are respectively \emph{$\Delta$-orthogonal} and \emph{$\Delta$-symmetric},\footnote{The names reflect the fact that for $\bDelta=\mathds{1}$, the concepts of `$\Delta$-transposition' and `$\Delta$-orthogonal' and `$\Delta$-symmetric' matrices reduce to the ordinary notion of `transposition' and `orthogonal' and `symmetric' matrices.}
\beq
\varphi_{\Delta}(\bO) = \bO^{-1} \textrm{~~~and~~~} \varphi_{\Delta}(\bR) = \bR\,.
\eeq
Since $\bU_{\textrm{m.c.}}^\zero(\bx)\in G$, it admits a decomposition as in eq.~\eqref{eq:OR},
\beq\label{eq:UOR}
\bU_{\textrm{m.c.}}^\zero(\bx) = \bO(\bx)\bR(\bx)\,.
\eeq
where we can explicitly write 
\begin{equation}
\bR(x)=\left(
\begin{array}{ccccccccccc}
1&0&0\\
\bs{s}(\bx)&\mathds{1}&0\\
s_{10}(\bx)&{\bs{\rho(\bs{s}(\bx))^T}}&1
\end{array}
\right) \textrm{~~~~and~~~~}\bs{O}(x)=\left(
\begin{array}{ccccccccccc}
1&0&0\\
\bs{G}(\bx)&\mathds{1}&0\\
\tilde{G}_{0}(\bx)&{-\bs{\rho}(\bs{G}(\bx))^T}&1
\end{array}
\right)\,.
\label{symorth}
\end{equation}
with
\beq
\bs{s}=(s_1,\dots,s_9)^T \textrm{~~~~and~~~~} \bs{G}=(G_1,\dots,G_9)^T\,.
\eeq
Here $s_1,\ldots,s_{10}$ and $G_1,\ldots,G_{9}$ denote the independent entries  of the $\Delta$-symmetric and $\Delta$-orthogonal parts, respectively. The explicit expressions in terms of the $\eps$-functions $t_{i,j}$ are given in the appendix~\ref{appexplsplit}.
The remaining entries are given by the functions
\begin{equation}
{\bs{\rho}(\bs{s})}=\frac{1}{3}\left(\begin{array}{l}
-2s_1+s_2+s_3+s_4\\
s_1-2s_2+s_3+s_4\\
s_1+s_2-2s_3+s_4\\
s_1+s_2+s_3-2s_4\\
6 \left(2s_5+2s_6+s_7-s_8-2s_9\right)\\
12 \left(s_5+4s_6-s_7-2s_8-s_9\right)\\
6 \left(s_5-2 \left(s_6-4s_7+s_8+2s_9\right)\right)\\
-6 \left(s_5+4s_6+2s_7-8s_8-4s_9\right)\\
-12 \left(s_5+s_6+2s_7-2 s_8-4 s_9\right)\end{array}\right)\,.
\end{equation}
and
\begin{align}
\tilde{G}_{0}&=\frac{1}{3} \Big[G_1^2+G_2^2+G_3^2+G_4^2-6 G_5^2-24( G_6^2+G_7^2+G_8^2+G_9^2)\\
&-\left(G_2+G_3+G_4\right) G_1-G_3 G_4-G_2 \left(G_3+G_4\right)
-6 \left(2 G_6 G_5+\left(G_5-2 G_6\right) G_7\right)\notag\\
&+6 \left(G_5+4 G_6+2 G_7\right) G_8+12 \left(G_5+G_6+2 G_7-2 G_8\right) G_9\Big]\,.\notag
\end{align}

The claim is that the $\Delta$-symmetric part $\bR(\bx)$ is always expressible in terms of rational functions, periods, and their derivatives, while the $\Delta$-orthogonal part is not constrained by eq.~\eqref{eq:DEQ_symmetry}. Moreover, the constraints that fix $\bR(\bx)$ can always be reduced to linear constraints. The decomposition of eq.~\eqref{eq:OR} therefore provides a parametrization of the matrix $\bU_{\textrm{m.c.}}^\zero(\bx)$ where the $\eps$-functions that can be expressed in terms of rational functions and periods via eq.~\eqref{eq:DEQ_symmetry} are made explicit. We stress that these statements are general and go beyond the case of the three-loop banana integral considered here. Rigorous proofs and additional examples will be provided in ref.~\cite{paper1}. In the remainder of this paper, we restrict ourselves to illustrating these general results on the concrete matrices for the three-loop banana integrals.

We see that $\bR(\bx)$ and $\bs{O}(\bx)$ are parametrized by 10 and 9 independent functions, respectively. If this decomposition is inserted into the symmetry constraint in eq.~\eqref{eq:DEQ_symmetry}, then the functions $G_i(\bx)$ appearing in $\bs{O}(\bx)$ drop out, while the functions $s_i(\bx)$ are fixed in terms of rational functions and periods by solving linear equations.
In our case, the $\Delta$-symmetric part of eq.~\eqref{symorth} is fixed such that
\begin{align}\label{eq:s_k}
s_k=\frac{-1}{2\bs{C}^{(0)}_{1,11}}\begin{cases}
\sum_{i=1}^4\bs{C}^\one_{i+1,11}\,\bs{j}_{i,k}\,,\quad&\text{if }k\in\{1,\dots,4\}\,,\\\\
\psi_0\bs{C}^\zero_{k,11}\,,\quad &\text{if }k\in\{5,\dots,10\}\,,
\end{cases}
\end{align}
and 
\begin{align}\label{eq:s_10}
 s_{10}=&\frac{\psi_0^2}{8(\bs{C}^\zero_{1,11})^2} \biggl[- 3 \sum_{i=2}^{5}(\bs{C}^\one_{i,11})^2 \,\bs{Z}_{i,i}  -6\sum_{i=2}^4\sum_{j=i+1}^5\bs{C}^\one_{i,11}\,\bs{C}^\one_{j,11} \,\bs{Z}_{i,j}\\
   &+
   12\,\bs{C}^\zero_{6,11}\,(\bs{C}^\zero_{6,11}+\bs{C}^\zero_{8,11}-\bs{C}^\zero_{9,11})-48\left(\bs{C}^\zero_{7,11}\,\bs{C}^\zero_{9,11}+\bs{C}^\zero_{10,11}\,(\bs{C}^\zero_{8,11}-\bs{C}^\zero_{9,11})\right) \nonumber \\
   &+
   24\left(\bs{C}^\zero_{6,11}\,(\bs{C}^\zero_{7,11}-\bs{C}^\zero_{10,11})-\bs{C}^\zero_{7,11}\,(\bs{C}^\zero_{8,11}-\bs{C}^\zero_{10,11})-\bs{C}^\zero_{8,11}\,\bs{C}^\zero_{9,11}\right)\nonumber\\
   &+  48 \sum_{i=7}^{10}\,(\bs{C}^\zero_{i,11})^2 -\bs{C}^{\scriptscriptstyle (2)}_{11,11}  \nonumber
  \biggr]\,.
\end{align}
The independent entries $G_1,\ldots, G_9$ of the $\Delta$-orthogonal part are not fixed.
They can be written down explicitly in terms of iterated integrals involving rational functions and periods and their derivatives (see section~\ref{sec:analytic}).
To conclude, we see that 10 of the 23 $\eps$-functions can be fixed, demonstrating the power of the method introduced here. The final differential equation matrix $\bA(\bx)$ with the expressions for the $s_i$ inserted and depending on the $\eps$-functions $G_i$ is given in \cite{bonndata}.

Let us conclude this discussion by making two comments. First, for our method to work, we require the matrix $\bDelta$. We determined $\bDelta$ from the intersection matrix $\bC(\bx,\eps)$ in the original basis by rotating it to the canonical basis, cf.~eq.~\eqref{eq:constant_intersection}. Given that it may not be easy to determine $\bC(\bx,\eps)$ in practice, one may ask how easily our method can be applied to other Feynman integrals. In ref.~\cite{paper1} we argue that it is often possible to determine $\bDelta$ directly without needing to know $\bC(\bx,\eps)$. We observe, however, that the knowledge of $\bC(\bx,\eps)$ has an advantage: We find that all the rational functions that appear in $\bR(\bx)$ can actually be written in very compact form in terms of the Laurent coefficients $\bC^{(i)}(\bx)$ of the entries of the intersection matrix $\bC(\bx,\eps)$, cf. eqs.~\eqref{eq:s_k} and~\eqref{eq:s_10}. Second, since our method does not allow us to constrain the functions $G_i(\bx)$ that appear in $\bO(\bx)$, it is natural to wonder if it is possible to constrain these functions by other means. While we do not have a definite answer to this question, we observe that in many cases one can indeed show that the entries in $\bO(\bx)$ cannot be expressed in terms of rational functions and (derivatives of) periods~\cite{paper1}. We therefore believe that the $\Delta$-orthogonal part $\bO(\bx)$ captures the genuinely new functions that go beyond rational functions, periods, and their derivatives. In particular, by comparing to the canonical basis for the equal-mass case, we can exclude that all the $G_i(\bx)$ in our case can be expressed in terms of rational functions and periods, and there must be at least one genuine $\eps$-function.

\section{A minimal representation for the $\eps$-functions}
\label{sec:analytic}
From the analysis in the previous section it follows that we can reduce the number of $\eps$-functions which cannot be expressed in terms of rational functions and (derivatives of) K3 periods from 23 to 13.
In this section we present analytic results in terms of iterated integrals for the remaining 13 $\eps$-functions. In our case, all $\eps$-functions depend on the four arguments $\bx$. Using symmetries related to permutations of the arguments, we manage to express all 13 $\eps$-functions in terms of only two new functions evaluated at permutations of their arguments.

\subsection{Analytic solution for the \texorpdfstring{$\eps$}{eps}-functions}
\label{solnewfunctions}

We start by presenting the explicit expressions for the $\eps$-functions that enter the differential equation matrix. Our $\eps$-functions satisfy first-order differential equations, which can be solved in terms of (iterated) integrals.
We present them by increasing level of complexity.

Let us start by discussing $G_9(\bx)$. The solution of the differential equation with respect to $x_1$ gives\footnote{In the remaining of this subsection, the variables inside the integral are omitted for clarity. See section~\ref{sec:minimal} for the full dependence.}
\begin{equation}
\label{solG_7}
   G_9(\bx)=
  \frac{\bs{C}^\zero_{10,11}(\bx)}{2\,\bs{C}^\zero_{1,11}(\bx)
  }\psi_0(\bx)+  \int_{\xi_1}^{x_1} \textrm{d}y_1\,
x_2\partial_{x_2}\psi_0(y_1,x_2,x_3,x_4)  + c_{G_9}\,,
\end{equation}
where $\bs{\xi}=(\xi_1,\ldots,\xi_4)$ is a generic point (we typically pick $\bs{\xi}=0$), and $c_{G_9}$ is an integration constant.
The previous expression is obtained by solving the differential equation in $x_1$, and so we expect the integration constant $c_{G_9}$ to be a function of the remaining variables $(x_2,x_3,x_4)$. After inserting these expressions into the differential equations with respect to $(x_2,x_3,x_4)$, we actually observe that $c_{G_9}$ is a constant, and the whole dependence on the other kinematic parameters is captured by eq.~\eqref{solG_7}.

We proceed in a similar fashion for the other functions. In all cases we observe that the differential equations with respect to a subset of the $x_i$ are sufficient to determine the function completely. We only present the minimal set of equations or iterated integrals, and the quantities $c_{G_i}$ are constants.
In this way we find for example:
\beq\bsp
    G_8(\bx)=& \frac{\bs{C}^\zero_{9,11}(\bx)}{2\,\bs{C}^\zero_{1,11(\bx)}
  }\psi_0(\bx) - 
  x_2 \psi_0(\bx)-  \int^{x_2}_{\xi_2} \textrm{d}y_2\,(x_4 
\partial_{x_4}\psi_0+ x_3 \partial_{x_3}\psi_0 + x_1
\partial_{x_1}\psi_0)\,+ c_{G_8}\,,\\
    G_7(\bx)=& \frac{\bs{C}^\zero_{8,11}(\bx)}{2\,\bs{C}^\zero_{1,11}(\bx)
  }\psi_0(\bx)+
  x_1 \psi_0(\bx)+  \int^{x_1}_{\xi_1} \textrm{d}y_1\,(x_4 
\partial_{x_4}\psi_0+ x_3 \partial_{x_3}\psi_0 + x_2
\partial_{x_2}\psi_0)\,+ c_{G_7}\,.
\esp\eeq
To fix $G_6(\bx)$ and $G_5(\bx)$ we need the solution of the differential equations with respect to $x_1$ and $x_3$. However, the differential equation with respect to $x_3$ is the same as the one with respect to $x_1$ just with $x_1 \leftrightarrow x_3$, so we get
\begin{align}
\nonumber G_6(\bx)\,&=\frac{\bs{C}^\zero_{7,11}(\bx)}{2\,\bs{C}^\zero_{1,11}(\bx)}\psi_0(\bx) + (x_1+x_3)\psi_0(\bx)+
    \int^{x_1}_{\xi_1} \textrm{d}y_1\,(x_4 
\partial_{x_4}\psi_0 \nonumber\\
&+ x_2
\partial_{x_2}\psi_0+ x_3 \partial_{x_3}\psi_0)+\int^{x_3}_{\xi_3} \textrm{d}y_3\,(x_4 
\partial_{x_4}\psi_0 + x_2
\partial_{x_2}\psi_0+ x_1 \partial_{x_1}\psi_0)+c_{G_6}\,,\\
\nonumber G_5(\bx)\,&= \frac{\bs{C}^\zero_{6,11}(\bx)}{2\,\bs{C}^\zero_{1,11}(\bx)}\psi_0(\bx)- (x_1+x_3)\psi_0(\bx)- \int_{\xi_1}^{x_1}\textrm{d}y_1\,x_3 \partial_{x_3}\psi_0-\int_{\xi_3}^{x_3}\textrm{d}y_3\,x_1 \partial_{x_1}\psi_0+ c_{G_5}\,.
\end{align}
Let us now look at $G_1(\bx)\,,G_2(\bx)\,,G_3(\bx)\,,G_4(\bx)$. 
From the series representation, we observe that it is sufficient to use the differential equation with respect to $x_i$ to determine $G_i(\bx)$, for $i=1,...,4\,$. After integration we get
\begin{align}
G_1(\bx)=&\sum_{i=1}^4\frac{\bs{C}^\one_{i+1,11}(\bx)\,\bs{j}_{i,1}(\bx)}{2\,\bs{C}^\zero_{1,11}(\bx)}-\int^{x_1}_{\xi_1}\,\mathrm{d}y_1\biggl\{-\bs{j}_{1,1}\frac{8\,\bs{C}^\zero_{1,11}\,G_0}{\psi_0^2}\nonumber\\
\label{eq:II_G1}&+\bs{j}_{2,1}\,h_1 +\bs{j}_{3,1}\,h_1[x_2\leftrightarrow x_3]+\bs{j}_{4,1}\,h_1[x_2\leftrightarrow x_4]\biggl\}+c_{G_1}\,,\\
    G_2(\bx)=&\sum_{i=1}^4\frac{\bs{C}^\one_{i+1,11}(\bx)\,\bs{j}_{i,2}(\bx)}{2\,\bs{C}^\zero_{1,11}(\bx)}-\int^{x_2}_{\xi_2}\,\mathrm{d}y_2\biggl\{\bs{j}_{2,2}\biggl[-\frac{8\,x_1\,\bs{C}^\zero_{1,11}\,G_0}{y_2\,\psi_0^2}+\frac{\partial_{x_1}\psi_0-\partial_{y_2}\psi_0}{y_2\,\psi_0}\biggr]\nonumber\\
    &+\bs{j}_{1,2}\,h_1+\bs{j}_{3,2}\,h_1[x_1\leftrightarrow x_3]+\bs{j}_{4,2}\,h_1[x_1 \leftrightarrow x_4]\biggl\}+c_{G_2}\,,\\
    G_3(\bx)=&\sum_{i=1}^4\frac{\bs{C}^\one_{i+1,11}(\bx)\,\bs{j}_{i,3}(\bx)}{2\,\bs{C}^\zero_{1,11}(\bx)}-\int^{x_3}_{\xi_3}\,\mathrm{d}y_3\biggl\{\bs{j}_{3,3}\biggl[-\frac{8\,x_1\,\bs{C}^\zero_{1,11}\,G_0}{y_3\,\psi_0^2}+\frac{\partial_{x_1}\psi_0-\partial_{y_3}\psi_0}{y_3\,\psi_0}\biggr]\nonumber\\
    &+\bs{j}_{1,3}\,h_1[x_2\leftrightarrow y_3]+\bs{j}_{2,3}\,h_1[x_1\leftrightarrow y_3]+\bs{j}_{4,3}\,h_1[x_2\leftrightarrow x_4,x_1\leftrightarrow y_3]\biggl\}+c_{G_3}\,,\\
    G_4(\bx)=&\sum_{i=1}^4\frac{\bs{C}^\one_{i+1,11}(\bx)\,\bs{j}_{i,4}(\bx)}{2\,\bs{C}^\zero_{1,11}(\bx)}-\int^{x_4}_{\xi_1}\,\mathrm{d}y_4\biggl\{\bs{j}_{4,4}\biggl[-\frac{8\,x_1\,\bs{C}^\zero_{1,11}\,G_0}{y_4\,\psi_0^2}+\frac{\partial_{x_1}\psi_0-\partial_{y_4}\psi_0}{y_4\,\psi_0}\biggr]\nonumber\\
\label{eq:II_G4}    &+\bs{j}_{1,4}\,h_1[x_2\leftrightarrow y_4]+\bs{j}_{2,4}\,h_1[x_1\leftrightarrow y_4]+\bs{j}_{3,4}\,h_1[x_2\leftrightarrow x_3,x_1\leftrightarrow y_4]\biggl\}+ c_{G_4}\,,
\end{align}
with
\begin{align}
    h_1(\bx)=&(\bOmega_{1}^{(1)}(\bx))_{7,5}+ 4\frac{ \bs{C}^\zero_{2,3}(\bx)}{\psi_0(\bx)^2}G_0(\bx)+\frac{1}{\psi_0(\bx)}\biggl[(\bOmega_1^{(1)}(\bx))_{7,9}\,\partial_{x_4}\psi_0(\bx) \nonumber\\
 &+(\bOmega_1^{(1)}(\bx))_{7,8}\,\partial_{x_3}\psi_0(\bx)+(\bOmega_1^{(1)}(\bx))_{7,7}\,\partial_{x_2}\psi_0(\bx) + (\bOmega_1^{(1)}(\bx))_{7,6} \,\partial_{x_1}\psi_0(\bx)\biggr]\,.
\end{align}

Let us now consider the differential equation for $G_0$, introduced in  eq.~\eqref{eq:Utm1}. As before, the series expansion tells us that it is enough to solve the differential equation with respect to $x_1$ and $x_2$, because the differential equations with respect to $x_3$ and $x_4$ can be obtained by exchanging $x_2\leftrightarrow x_3$ and $x_2\leftrightarrow x_4$. These differential equations read
\begin{align}
    -\partial_{x_1}G_0\,&=r_0\,\psi_0^2+r_1\,\psi_0\,\partial_{x_1}\psi_0+r_2\,\psi_0\,\partial_{x_2}\psi_0+r_2[x_2\leftrightarrow x_3]\,\psi_0\,\partial_{x_3}\psi_0\nonumber\\
    &+r_2[x_2\leftrightarrow x_4]\,\psi_0\,\partial_{x_4}\psi_0+r_3\,(\partial_{x_1}\psi_0)\,(\partial_{x_2}\psi_0)\nonumber\\
    &+r_3[x_2\leftrightarrow x_3]\,(\partial_{x_1}\psi_0)\,(\partial_{x_3}\psi_0)+r_3[x_2\leftrightarrow x_4]\,(\partial_{x_1}\psi_0)\,(\partial_{x_4}\psi_0)\nonumber\\
    &+r_4\,(\partial_{x_2}\psi_0)\,(\partial_{x_3}\psi_0)+r_4[x_3\leftrightarrow x_4]\,(\partial_{x_2}\psi_0)\,(\partial_{x_4}\psi_0)\nonumber\\
    &+r_4[x_2\leftrightarrow x_4]\,(\partial_{x_4}\psi_0)\,(\partial_{x_3}\psi_0)\,,\\
    -\partial_{x_2}G_0&\,=r_5\,\psi_0^2+r_6\,\psi_0\,\partial_{x_1}\psi_0+r_7\,\psi_0\,\partial_{x_2}\psi_0+r_8\,\psi_0\,\partial_{x_3}\psi_0\nonumber\\
    &+r_8[x_3\leftrightarrow x_4]\,\psi_0\,\partial_{x_4}\psi_0+r_9\,(\partial_{x_1}\psi_0)\,(\partial_{x_2}\psi_0)+r_{10}\,(\partial_{x_1}\psi_0)\,(\partial_{x_3}\psi_0)\nonumber\\
    &+r_{10}[x_3\leftrightarrow x_4]\,(\partial_{x_1}\psi_0)\,(\partial_{x_4}\psi_0)+r_{11}\,(\partial_{x_2}\psi_0)\,(\partial_{x_3}\psi_0)\nonumber\\
    &+r_{11}[x_3\leftrightarrow x_4]\,(\partial_{x_2}\psi_0)\,(\partial_{x_4}\psi_0)+r_{12}\,(\partial_{x_4}\psi_0)\,(\partial_{x_3}\psi_0)\,,\nonumber
\end{align}    
where $r_i$ are rational functions in the kinematics and are available in in computer readable form~\cite{bonndata}. We note that we can also use eqs.~(\ref{eq:II_G1} - \ref{eq:II_G4}) to express $G_0(\bx)$ in terms of the derivative $\partial_{x_i}G_i(\bx)$.

Let us now solve for the $\eps$-functions associated with the tadpole subsector. Looking at the series representations, we see that
\beq\bsp
    G_{11}(\bx)=G_{10}[x_1\leftrightarrow x_2]\,,\\
    G_{12}(\bx)=G_{10}[x_1\leftrightarrow x_3]\,,\\
    G_{13}(\bx)=G_{11}[x_2\leftrightarrow x_4]\,.
\esp\eeq
Hence, we just need to solve for $G_{10}(\bx)\,$. From the series representation we see that to fully fix $G_{10}(\bx)$, we need the differential equations with respect to $x_1\,,x_2\,,x_3\,$ (the differential equation with respect to $x_4$ is the same as with respect to $x_3$ with $x_3\leftrightarrow x_4\,$). We get
\begin{align}
    G_{10}=&-((\bOmega_1^{(2)}(\bx))_{7,1} \bZ^{-1}_{2,3}(\bx) + 
  (\bOmega_1^{(2)}(\bx))_{8,1} \bZ^{-1}_{2,4}(\bx) + 
  (\bOmega_1^{(2)}(\bx))_{9,1}\bZ^{-1}_{2,5}(\bx))\psi_0(\bx)\\
&-4
    \int^{x_1}_{\xi_1} \textrm{d}y_1\,\left(x_4 
\partial_{x_4}\psi_0 + x_3 \partial_{x_3}\psi_0\right)-2\int^{x_2}_{\xi_2} \textrm{d}y_2\,\left(x_4 
\partial_{x_4}\psi_0 + x_3
\partial_{x_3}\psi_0+ x_1 \partial_{x_1}\psi_0\right)\nonumber\\
&+\frac{2(1-x_1-3\,x_2)}{3}\psi_0(\bx)+c_{G_{10}}\,.\nonumber
\end{align}

\subsection{A minimal set of (iterated) integrals}
\label{sec:minimal}
All the $\eps$-functions defined in the previous subsection are functions of the four variables $\bx$. The banana integrals come with a natural action of the permutation group that exchanges the four masses, or equivalently the entries of $\bx$. From their series representations, we see that all the 13 independent $\eps$-functions defined in the previous subsection can be expressed in terms of just two new functions evaluated at permuted arguments. Note that in order to make the permutation symmetries manifest, we need to pick the boundary conditions appropriately. In the following, we pick the totally symmetric point $\bs{\xi}=0$ and we fix the integration constants $c_{G_i}= 0$ . 

Let us define
\begin{align}
\mathcal{I}_1(x_i,x_j,x_k,x_l)=&\int_{0}^{x_i} \textrm{d}y_i\,
x_j\partial_{x_j}\psi_0(y_i,x_j,x_k,x_l)\,,\\
\mathcal{I}_2(x_i,x_j,x_k,x_l)=&\int^{x_i}_{0}\,\mathrm{d}y_i\biggl[-\bs{j}_{1,1}(y_i,x_j,x_k,x_l)\frac{8\,\,\bs{C}^\zero_{1,11}(y_i,x_j,x_k,x_l)\,G_0(y_i,x_j,x_k,x_l)}{\,\psi_0(y_i,x_j,x_k,x_l)^2}\nonumber\\
&+\sum_{\sigma}\bs{j}_{2,1}(y_i,x_{\sigma{(j)}},x_{\sigma{(k)}},x_{\sigma{(l)}})\,h_1(y_i,x_{\sigma{(j)}},x_{\sigma{(k)}},x_{\sigma{(l)}}) \biggr]\,,
\end{align}
where we sum over the permutations $\sigma\in \{e,(j\,k),(j\,l)\}$.
Note that the integrand defining $\mathcal{I}_2(\bx)$ depends on $G_0(\bx)$, which is itself defined by an integral. Hence, $\mathcal{I}_2(\bx)$ is in fact an iterated integral of length two. We also introduce the shorthand
\beq
\mathcal{C}^{(j)}_i(\bx):= \frac{\bs{C}^{(j)}_{i+1,11}(\bx)}{2\,\bs{C}^\zero_{1,11}(\bx)
  }\psi_0(\bx)\,.
  \eeq

We can now observe that all 13 $\eps$-functions from the previous subsection can be expressed in terms of the two functions just defined. We have
\begin{align}
G_1(\bs{x})=\,&\sum_{k=1}^4\mathcal{C}^\one_1(\bs{x})\frac{\bs{j}_{k,1}(\bs{x})}{\psi_0(\bs{x})}-\mathcal{I}_2(x_1,x_2,x_3,x_4)\,,\nonumber\\
G_2(\bs{x})=\,&\sum_{k=1}^4\mathcal{C}^\one_2(\bs{x})\frac{\bs{j}_{k,2}(\bs{x})}{\psi_0(\bs{x})}-\mathcal{I}_2(x_2,x_1,x_3,x_4)\,,\nonumber\\
G_3(\bs{x})=\,&\sum_{k=1}^4\mathcal{C}^\one_3(\bs{x})\frac{\bs{j}_{k,3}(\bs{x})}{\psi_0(\bs{x})}-\mathcal{I}_2(x_3,x_2,x_1,x_4)\,,\nonumber\\
G_4(\bs{x})=\,&\sum_{k=1}^4\mathcal{C}^\one_4(\bs{x})\frac{\bs{j}_{k,4}(\bs{x})}{\psi_0(\bs{x})}-\mathcal{I}_2(x_4,x_2,x_3,x_1)\,,\nonumber\\
G_5(\bs{x})=\,& \mathcal{C}^\zero_5(\bs{x})- (x_1+x_3)\psi_0(\bs{x})- \mathcal{I}_1(x_1,x_3,x_2,x_4)-\mathcal{I}_1(x_3,x_1,x_2,x_4)\,,\nonumber\\
G_6(\bs{x})=\,&\mathcal{C}^\zero_6(\bs{x})+ (x_1+x_3)\psi_0(\bs{x})+\mathcal{I}_1(x_1,x_4,x_2,x_3)
+\mathcal{I}_1(x_1,x_2,x_3,x_4)\nonumber\\
&+\mathcal{I}_1(x_1,x_3,x_2,x_4)+\mathcal{I}_1(x_3,x_4,x_1,x_2)+\mathcal{I}_1(x_3,x_2,x_1,x_4)+\mathcal{I}_1(x_3,x_1,x_2,x_4)\,,\nonumber\\
 \label{eq:G_i_final}   G_7(\bs{x})=\,&\mathcal{C}^\zero_7(\bs{x})+
  x_1 \psi_0(\bs{x})+ \mathcal{I}_1(x_1,x_4,x_2,x_3)+ \mathcal{I}_1(x_1,x_3,x_2,x_4)+ \mathcal{I}_1(x_1,x_2,x_3,x_4)\,,\\
  G_8(\bs{x})=\,&\mathcal{C}^\zero_8(\bs{x})- 
  x_2 \psi_0(\bs{x})- \mathcal{I}_1(x_2,x_4,x_1,x_3)-\mathcal{I}_1(x_2,x_3,x_1,x_4)-\mathcal{I}_1(x_2,x_1,x_3,x_4)\,,\nonumber\\
   G_9(\bs{x})=\,&
  \mathcal{C}^\zero_9(\bs{x})+  \mathcal{I}_1(x_1,x_2,x_3,x_4)\,,\nonumber\\
  G_{10}(\bs{x})=\,&\biggl[\frac{2(1-x_1-3\,x_2)}{3}-(\bOmega_{7,1}^{(2)}(\bx) \bZ^{-1}_{2,3}(\bx) + 
  \bOmega_{8,1}^{(2)}(\bx) \bZ^{-1}_{2,4}(\bx) \nonumber\\
&+ 
  \bOmega_{9,1}^{(2)}(\bx) \bZ^{-1}_{2,5}(\bx))\biggr]\psi_0(\bs{x})-4\,\mathcal{I}_1(x_1,x_4,x_2,x_3)-4\,\mathcal{I}_1(x_1,x_3,x_2,x_4)\nonumber\\
&-2\,\mathcal{I}_1(x_2,x_4,x_1,x_3)-2\,\mathcal{I}_1(x_2,x_3,x_1,x_4)-2\,\mathcal{I}_1(x_2,x_1,x_3,x_4)\,,\nonumber\\
    G_{11}(\bs{x})=\,&G_{10}(x_2,x_1,x_3,x_4)\,,\nonumber\\
    G_{12}(\bs{x})=\,&G_{10}(x_3,x_2,x_1,x_4)\,,\nonumber\\
    G_{13}(\bs{x})=\,&G_{11}(x_1,x_4,x_3,x_2)\,.\nonumber
\end{align}
These expressions are our final result for the $\eps$-functions that appear in the canonical differential equation for the banana integrals. We find it remarkable that only two genuinely new functions (evaluated at permuted arguments) are required. The expressions in eq.~\eqref{eq:G_i_final} are our final results for the $\eps$-functions. The final form of the differential equation matrix in canonical form is provided in computer readable form~\cite{bonndata}.

\section{Conclusions and outlook}
\label{sec:conclusions}

In this paper we have obtained for the first time a system of canonical differential equations satisfied by the master integrals of the three-loop banana integral family with four distinct non-zero masses in $D=2-2\eps$ dimensions. Combined with the knowledge of the small-mass limit of the banana integrals (see ref.~\cite{Bonisch:2021yfw} and appendix~\ref{app:init_cond}), this allows us to obtain analytic results for all master integrals in terms of iterated integrals over a set of integration kernels that involve rational functions, (derivatives of) periods of the underlying family of K3 surface, as well as integrals involving these functions. Our final result for the differential equation is lengthy, and it is provided in computer readable form~\cite{bonndata}.

Our result was made possible by combining tools from three different advances made over the last few years. In particular, we apply the method of refs.~\cite{Gorges:2023zgv,Duhr:2025lbz,Maggio:2025jel} to construct a sequence of transformations that brings the system into an $\eps$-factorized form. We observe that the resulting $\eps$-factorized system has additional properties, which confirm those expected from systems in canonical form in ref.~\cite{Duhr:2025lbz}. An important ingredient in order to construct the transformation matrix using the method of refs.~\cite{Gorges:2023zgv,Duhr:2025lbz,Maggio:2025jel} is an understanding of the geometry of the family of K3 surfaces associated to the three-loop banana integrals, in particular their periods and the quadratic relations they satisfy. Finally, the method of refs.~\cite{Gorges:2023zgv,Duhr:2025lbz,Maggio:2025jel} introduces 23 functions -- which we called $\eps$-functions~-- defined as iterated integrals over rational functions, K3 periods, and their derivatives. We used tools from twisted cohomology to derive relations between these functions. This allowed us to reduce the number of $\eps$-functions from 23 to 13. Once symmetries are taken into account this number is further reduced to two. 

We expect that the techniques we have applied here to derive relations between $\eps$-functions, in particular the techniques from twisted cohomology, will also be useful to simplify the differential equations for other classes of Feynman integrals. We leave the presentation of additional examples, as well as a further mathematical derivations, for future work~\cite{paper1}.

\acknowledgments

We are grateful to Gaia Fontana, Christoph Nega, Lorenzo Tancredi, Sid Smith, Yoann Sohnle and Federico Gasparotto for discussions.
This work is supported in part by the CRC 1639 ``NuMeriQS'' (CS) and by the European Union
(ERC Consolidator Grant LoCoMotive 101043686) (CD, FP, SM, SFS). Views
and opinions expressed are however those of the author(s)
only and do not necessarily reflect those of the European
Union or the European Research Council. Neither the
European Union nor the granting authority can be held
responsible for them.

\begin{appendix}
\section{Decomposition into Orthogonal and Symmetric Part}
\label{appexplsplit}
Here we present the expressions of $s_k$ and $G_i$ that appear in the decomposition of $\bU^{(0)}_{\textrm{t}}$ into $\Delta$-symmetric and $\Delta$-orthogonal parts, cf. eq.~\eqref{eq:UOR}, in terms of the $\eps$-functions $t_{i,j}$. 
For the independent entries $s_k$ of the $\Delta$-symmetric part, we have
\begin{align}
\bs{s}&=\frac{1}{2}\left(\begin{array}{l}
t_{6,5}+t_{15,7}+t_{15,8}+t_{15,9}\\
t_{7,5}+t_{15,6}+t_{15,8}+t_{15,9}\\
t_{8,5}+t_{15,6}+t_{15,7}+t_{15,9}\\
t_{9,5}+t_{15,6}+t_{15,7}+t_{15,8}\\
\frac{1}{12} \left(12 t_{10,5}+6 t_{15,10}-2 t_{15,11}-t_{15,12}-t_{15,13}+t_{15,14}\right)\\
\frac{1}{12} \left(12 t_{11,5}-2 t_{15,10}+2 t_{15,11}+t_{15,12}+t_{15,13}\right)\\
\frac{1}{24} \left(24 t_{12,5}-2 t_{15,10}+2 t_{15,11}+3 t_{15,12}+t_{15,13}+t_{15,14}\right)\\
\frac{1}{24} \left(24 t_{13,5}-2 t_{15,10}+2 t_{15,11}+t_{15,12}+3 t_{15,13}-t_{15,14}\right)\\
\frac{1}{24} \left(24 t_{14,5}+2 t_{15,10}+t_{15,12}-t_{15,13}+3 t_{15,14}\right)
\end{array}
\right) \,,
\end{align}
and
\begin{align}
s_{10}&=\frac{1}{192} \Big\{288 t_{10,5}^2+1152( t_{11,5}^2+ t_{12,5}^2+t_{13,5}^2+ t_{14,5}^2- t_{12,5} t_{14,5}+ t_{13,5} t_{14,5})\notag\\
&-576 t_{12,5} t_{13,5}+192 t_{15,5}+48\Big[-t_{6,5}^2+\left(t_{7,5}+t_{8,5}+t_{9,5}-t_{15,6}\right) t_{6,5}\notag\\
&-t_{7,5}^2-t_{8,5}^2-t_{9,5}^2+t_{8,5} t_{9,5}+t_{7,5} \left(t_{8,5}+t_{9,5}-t_{15,7}\right)-t_{15,6} t_{15,7}\notag\\
&-\left(t_{8,5}+t_{15,6}+t_{15,7}\right) t_{15,8}-\left(t_{9,5}+t_{15,6}+t_{15,7}+t_{15,8}\right) t_{15,9}\Big]\notag\\
&+48 t_{10,5} \left(12 t_{11,5}+6 t_{12,5}-6 \left(t_{13,5}+2 t_{14,5}\right)-t_{15,10}\right)-12 t_{15,10}^2+8 t_{15,11} t_{15,10}\notag\\
&-48 t_{11,5} \left(12 \left(t_{12,5}+2 t_{13,5}+t_{14,5}\right)+t_{15,11}\right)-48 (t_{12,5} t_{15,12}+ t_{13,5} t_{15,13})\notag\\
&-2 t_{15,13} t_{15,12}-3( t_{15,13}^2+t_{15,14}^2+ t_{15,12}^2)+4 t_{15,12} t_{15,10}\notag\\
&+4 (- t_{15,11}^2+t_{15,10} t_{15,13}- t_{15,11} t_{15,13}- t_{15,11} t_{15,12})\notag\\
&-2 \left(24 t_{14,5}+2 t_{15,10}+t_{15,12}-t_{15,13}\right) t_{15,14}\Big\}
\notag \,.
\end{align}
For the independent entries $G_k$ of the $\Delta$-orthogonal part, we have
\begin{align}
\bs{G}&=\frac{1}{2}\left(\begin{array}{l}
t_{6,5}-t_{15,7}-t_{15,8}-t_{15,9}\\
t_{7,5}-t_{15,6}-t_{15,8}-t_{15,9}\\
t_{8,5}-t_{15,6}-t_{15,7}-t_{15,9}\\
t_{9,5}-t_{15,6}-t_{15,7}-t_{15,8}\\
\frac{1}{12} \left(12 t_{10,5}-6 t_{15,10}+2 t_{15,11}+t_{15,12}+t_{15,13}-t_{15,14}\right)\\
\frac{1}{12} \left(12 t_{11,5}+2 t_{15,10}-2 t_{15,11}-t_{15,12}-t_{15,13}\right)\\
\frac{1}{24} \left(24 t_{12,5}+2 t_{15,10}-2 t_{15,11}-3 t_{15,12}-t_{15,13}-t_{15,14}\right)\\
\frac{1}{24} \left(24 t_{13,5}+2 t_{15,10}-2 t_{15,11}-t_{15,12}-3 t_{15,13}+t_{15,14}\right)\\
\frac{1}{24} \left(24 t_{14,5}-2 t_{15,10}-t_{15,12}+t_{15,13}-3 t_{15,14}\right)\end{array}\right)\,.
\end{align}

\section{Expansion of the \texorpdfstring{$\eps$}{eps}-functions around the MUM-point}
\label{app:series}
In this appendix we present the first few terms in the expansion around the MUM-point $\bx=0$ of all $\eps$-functions $t_{i,j}$ that appear in $\bU_{\!\textrm{t}}(\bx)$. We show terms that are at most quadratic in the $x_i$, and the dots indicate contributions at least cubic in $x_i$. Here $c_{i,j}$ are integration constants.
\beq\bsp
    G_0(\bx)=\,& -\frac{1}{2}+\frac{1}{6} \, (3 x_1 + 4 (x_2 + x_3 + x_4)) + 
 \frac{1}{6}  \,(-3 x_1^2 - 4 x_2^2- 4 x_3^2 + 10 x_3 x_4 \\
 &- 4 x_4^2 + 
    10 x_2 (x_3 + x_4) + 5 x_1 (x_2 + x_3 + x_4))+\ldots\,,\\
t_{6,5}(\bx)=\,&c_{6,5} -6 \, x_1 - 2\,  x_1 (3 x_1 + 5 (x_2 + x_3 + x_4))+\ldots\,,\\
t_{7,5}(\bx)=\,&c_{7,5} -6\,  x_2 - 2\, x_2 (5 x_1 + 3 x_2 + 5 (x_3 + x_4))+\ldots\,,\\
t_{8,5}(\bx)=\,&c_{8,5}-6 \, x_3 - 2 \, x_3 (5 x_1 + 5 x_2 + 3 x_3 + 5 x_4)+\ldots\,,\\
t_{9,5}(\bx)=\,&c_{9,5}-6 \, x_4 - 2 \, x_4 (5 x_1 + 5 x_2 + 5 x_3 + 3 x_4)+\ldots\,,\\
t_{10,5}(\bx)=\,& c_{10,5}-(x_1 + x_3) - 
 (x_1^2 + x_3 (x_2 + x_3 + x_4) \\
 &+x_1 (x_2 + 4 x_3 + x_4))+\ldots\,,\\
 t_{11,5}(\bx)=\,&c_{11,5}+ (x_1 + x_3) + 
 (x_1^2 + 2 x_1 (x_2 + 2 x_3 + x_4) \\
 &+ x_3 (2 x_2 + x_3 + 2 x_4))+\ldots\,,\\
 t_{12,5}(\bx)=\,&c_{12,5}+ x_1 + x_1 (x_1 + 2 (x_2 + x_3 + x_4))+\ldots\,,\\
 t_{13,5}(\bx)=\,&c_{13,5}- x_2 -  x_2 (2 x_1 + x_2 + 2 (x_3 + x_4))+\ldots\,,\\
t_{14,5}(\bx)=\,&c_{14,5}+ x_1 x_2+\ldots\,,\\
t_{15,1}(\bx)=\,&c_{15,1}+\frac{1}{3}\,  (3 x_1 - x_2 - x_3 - x_4) + 
 \frac{1}{9}  \,(-9 x_1^2 - x_1 x_2 + 3 x_2^2 + 3 x_3^2 \\
 &- 55 x_3 x_4 + 
    3 x_4^2 + 41 x_1 (x_3 + x_4) - 13 x_2 (x_3 + x_4))+\ldots\,,\\
t_{15,2}(\bx)=\,&c_{15,2}+\frac{1}{3}\,  (-x_1 + 3 x_2 - x_3 - x_4) + 
 \frac{1}{9}\, (3 x_1^2 - 9 x_2^2 + 3 x_3^2 - 55 x_3 x_4 \\
 &+ 3 x_4^2 + 
    41 x_2 (x_3 + x_4) - x_1 (x_2 + 13 (x_3 + x_4)))+\ldots\,,\\
t_{15,3}(\bx)=\,&c_{15,3}+\frac{1}{3} \, (-x_1 - x_2 + 3 x_3 - x_4) + 
 \frac{1}{9}\,  (3 x_1^2 + 3 x_2^2 - 9 x_3^2\\
 & + 
    x_1 (-13 x_2 + 41 x_3 - 55 x_4) + 41 x_3 x_4 + 3 x_4^2 - 
    x_2 (x_3 + 13 x_4))+\ldots\,,\\
t_{15,4}(\bx)=\,&c_{15,4}+\frac{1}{3}\, (-x_1 - x_2 - x_3 + 3 x_4) + 
 \frac{1}{9}\, (3 x_1^2 + 3 x_2^2 - 55 x_2 x_3 \\
 &+ 3 x_3^2 + 41 x_2 x_4 + 
    41 x_3 x_4 - 9 x_4^2 - x_1 (13 x_2 + 13 x_3 + x_4))+\ldots\,,
\esp\eeq
From the relations in eqs.~\eqref{4massesrel1},~\eqref{4massesrel4},~\eqref{symorth}, and choosing the integration constants to be zero, we find the remaining functions 
\begin{align}
   \nonumber t_{15,14}(\bx)=\,&-4+\frac{28}{3} \, (x_1 x_2 - x_2 x_3 - x_1 x_4 + x_3 x_4)+\ldots\,,\\
 \nonumber   t_{15,13}(\bx)=\,&-8-4\, (x_2 - x_4) - 
 \frac{4}{3}  (10 x_1 x_2 - 3 x_2^2 + 17 x_2 x_3 - 10 x_1 x_4 \\
 \nonumber&- 
    17 x_3 x_4 + 3 x_4^2)+\ldots\,,\\
 \nonumber   t_{15,12}(\bx)=\,&2+4\,  (x_1 - x_3) - 
 \frac{4}{3} \, (3 x_1^2 - 10 x_1 x_2 + 10 x_2 x_3 - 3 x_3^2 - 17 x_1 x_4\\
\nonumber & + 
    17 x_3 x_4)+\ldots\,,
        \end{align}
    \begin{align}
\nonumber    t_{15,11}(\bx)=\,&8+4\,  (x_3 - x_4) + 
 \frac{4}{3}\,  (10 x_1 x_3 + 17 x_2 x_3 - 3 x_3^2 - 10 x_1 x_4 \\
\nonumber &- 
    17 x_2 x_4 + 3 x_4^2)+\ldots\,,\\
    t_{15,10}(\bx)=\,&2-\frac{14}{3}\,  (x_1 x_3 - x_2 x_3 - x_1 x_4 + x_2 x_4)+\ldots\,,\\
 \nonumber   t_{15,9}(\bx)=\,&-\frac{1}{6}+\frac{1}{6} (-8 x_1 - 5 (x_2 + x_3 - 2 x_4))+\frac{1}{18} (24 x_1^2 + x_1 (-227 x_2 - 227 x_3 \\
 \nonumber   &+ 76 x_4) + 
   5 (3 x_2^2 + 3 x_3^2 + 17 x_3 x_4 - 6 x_4^2 + 17 x_2 (-2 x_3 + x_4)))+\ldots\,,\\
  \nonumber  t_{15,8}(\bx)=\,&-\frac{1}{6}+\frac{1}{6} (-8 x_1 - 5 (x_2 - 2 x_3 + x_4))+\frac{1}{18} (24 x_1^2 + x_1 (-227 x_2 + 76 x_3 \\
\nonumber    &- 227 x_4) + 
   5 (3 x_2^2 - 6 x_3^2 + 17 x_2 (x_3 - 2 x_4) + 17 x_3 x_4 + 3 x_4^2))+\ldots\,,\\
 \nonumber   t_{15,7}(\bx)=\,&-\frac{1}{6}+\frac{1}{6} (-8 x_1 + 10 x_2 - 5 (x_3 + x_4))+\frac{1}{18} (24 x_1^2 + x_1 (76 x_2 - 227 (x_3 \\
 \nonumber   &+ x_4)) + 
   5 (-6 x_2^2 + 3 x_3^2 - 34 x_3 x_4 + 3 x_4^2 + 17 x_2 (x_3 + x_4)))+\ldots\,,\\
\nonumber    t_{15,6}(\bx)=\,&\frac{1}{12\,x_1} + \frac{-7  x_1 - 
  3 (x_2 + x_3 + x_4)}{12 x_1} \\
\nonumber  &+ 
  \frac{42 x_1^2 + 6 x_2^2 + 6 x_3^2 - 10 x_3 x_4 + 6 x_4^2 - 
    10 x_2 (x_3 + x_4) - 25 x_1 (x_2 + x_3 + x_4)}{18 x_1} \\
\nonumber  &+ 
 \frac{1}{18 x_1}
 (-42 x_1^3 - 6 x_2^3 - 6 x_3^3 + 20 x_3^2 x_4 + 
    20 x_3 x_4^2 - 6 x_4^3 \\
 \nonumber &+ 20 x_2^2 (x_3 + x_4) + 
    119 x_1^2 (x_2 + x_3 + x_4) 
    2 x_2 (10 x_3^2 - 63 x_3 x_4 + 10 x_4^2)\\
\nonumber  & + 
    5 x_1 (7 x_2^2 + 7 x_3^2 - 60 x_3 x_4 + 7 x_4^2 - 
       60 x_2 (x_3 + x_4)))+\ldots\,,\\
 \nonumber   t_{15,5}(\bx)=\,&2 + 2\,  (5 x_1 + 5 x_2 + 4 x_3 - x_4) + 
  \frac{2}{9} (9 x_1^2 + 9 x_2^2 + 221 x_2 x_3 + 38 x_2 x_4 \\
\nonumber &+ 
    29 x_3 x_4 - 27 x_4^2 + x_1 (200 x_2 + 191 x_3 + 68 x_4))+\ldots\,.
\end{align}
Inserting these expressions into the transformed differential equation, it is possible to check that it is $\eps$-factorized and only has simple poles (around the MUM-point $\bx=0$) in the kinematics.

We also present the expansions around the MUM-point of the functions $G_i(\bx)$ that appear in the $\Delta$-orthogonal part or the tadpole subsectors.
\begin{align}
    G_9(\bx)=\,&\frac{1}{72} (-3 - 6  (x_1 + x_2 - x_3 - x_4) + 
 2  (3 x_1^2 - 23 x_1 x_2 + 3 x_2^2 - 3 x_3^2 + 13 x_3 x_4 \nonumber\\
 &- 
    3 x_4^2 + 7 x_1 (x_3 + x_4) + 7 x_2 (x_3 + x_4)))+\ldots\,,\\
    G_8(\bx)=\,&\frac{1}{72}(9 - 6  (x_1 + 3 x_2 + x_3 + x_4) + 
 6  (x_1^2 - 9 x_2^2 + x_3^2 - 9 x_3 x_4 + x_4^2 \nonumber\\
 &- 
    3 x_2 (x_3 + x_4) - 3 x_1 (x_2 + 3 (x_3 + x_4))))+\ldots\,,\\
    G_7(\bx)=\,&\frac{1}{72}(-9 + 6  (3 x_1 + x_2 + x_3 + x_4) + 
 6  (9 x_1^2 - x_2^2 - x_3^2 + 9 x_3 x_4 - x_4^2 \nonumber\\
 &+ 
    9 x_2 (x_3 + x_4) + 3 x_1 (x_2 + x_3 + x_4)))+\ldots\,,\\
    G_6(\bx)=\,&\frac{1}{36}(-9 + 6  (2 x_1 + x_2 + 2 x_3 + x_4) + 
 6  (4 x_1^2 - x_2^2 + 6 x_2 x_3 + 4 x_3^2 + 9 x_2 x_4 \nonumber\\
 &+ 
    6 x_3 x_4 - x_4^2 + 3 x_1 (2 x_2 + x_3 + 2 x_4)))+\ldots\,,
        \end{align}
    \begin{align}
    G_5(\bx)=\,&\frac{1}{36}(3 - 6  (2 x_1 + x_2 + 2 x_3 + x_4) - 
 2  (12 x_1^2 - 3 x_2^2 + 16 x_2 x_3 + 12 x_3^2 + 13 x_2 x_4 \nonumber\\
 &+ 
    16 x_3 x_4 - 3 x_4^2 + x_1 (16 x_2 - 5 x_3 + 16 x_4)))+\ldots\,,\\
    G_1(\bx)=\,&\frac{1}{12} (3 - 12  x_1 + 6  x_1 (-10 x_1 + 11 (x_2 + x_3 + x_4))+\ldots\,,\\
    G_2(\bx)=\,&-\frac{1}{24 \, x_1} +\frac{ 
 11 x_1 + 3 (x_2 + x_3 + x_4)}{24 x_1} \nonumber\\
 &+ 
  \frac{6 (x_1^2-x_2^2-x_3^2-x_4^2) - 53 x_1 x_2  + 10 x_3 x_4  + 10 x_1 (x_3 + x_4) + 10 x_2 (x_3 + x_4)}{36 x_1} \nonumber\\
 &+ 
 \frac{1}{36 x_1}
   (-6 x_1^3+6 x_2^3 + 6 x_3^3  - 20 x_3^2 x_4 - 20 x_3 x_4^2 + 
    6 x_4^3 - 20 x_2^2 (x_3 + x_4)\nonumber\\
 & - 
    2 x_2 (10 x_3^2 - 63 x_3 x_4 + 10 x_4^2) + 
    x_1^2 (155 x_2 + 32 (x_3 + x_4)) \nonumber\\
 &- 
    x_1 (173 x_2^2 - 205 x_2 (x_3 + x_4) + 
       10 (2 x_3^2 - 13 x_3 x_4 + 2 x_4^2)))+\ldots\,,\\
    G_3(\bx)=\,&-\frac{1}{24 \, x_1} + \frac{11 x_1 + 3 (x_2 + x_3 + x_4)}{24 x_1} \nonumber\\
 &+ 
  \frac{6 (x_1^2-x_2^2-x_3^2-x_4^2) + 10 x_3 x_4 + 
    10 x_2 (x_3 + x_4) + x_1 (10 x_2 - 53 x_3 + 10 x_4)}{36 x_1} \nonumber\\
 &+ 
 \frac{1}{36 x_1}
   (-6 x_1^3 + 6 x_2^3 + 6 x_3^3 - 20 x_3^2 x_4 - 20 x_3 x_4^2 + 
    6 x_4^3 - 20 x_2^2 (x_3 + x_4)\nonumber\\
 & + 
    x_1^2 (32 x_2 + 155 x_3 + 32 x_4) - 
    2 x_2 (10 x_3^2 - 63 x_3 x_4 + 10 x_4^2) \nonumber\\
 &- 
    x_1 (20 x_2^2 + 173 x_3^2 - 205 x_3 x_4 + 20 x_4^2 - 
       5 x_2 (41 x_3 + 26 x_4)))+\ldots\,,\\
    G_4(\bx)=\,&-\frac{1}{24 \, x_1} +\frac{
 11 x_1 + 3 (x_2 + x_3 + x_4)}{24 x_1}\nonumber\\
 &+ 
  \frac{6 (x_1^2-x_2^2-x_3^2-x_4^2) + x_1 (10 x_2 + 10 x_3 - 53 x_4) + 
    10 x_3 x_4 + 10 x_2 (x_3 + x_4)}{36 x_1}\nonumber\\
 &  + 
 \frac{1}{36 x_1}
   (-6 x_1^3 + 6 x_2^3 + 6 x_3^3 - 20 x_3^2 x_4 - 20 x_3 x_4^2 + 
    6 x_4^3 - 20 x_2^2 (x_3 + x_4) \nonumber\\
 &+ 
    x_1^2 (32 x_2 + 32 x_3 + 155 x_4) - 
    2 x_2 (10 x_3^2 - 63 x_3 x_4 + 10 x_4^2) \nonumber\\
 &- 
    x_1 (20 x_2^2 + 20 x_3^2 - 205 x_3 x_4 + 173 x_4^2 - 
       5 x_2 (26 x_3 + 41 x_4)))+\ldots\,,\\
    G_0(\bx)=\,& -\frac{1}{2}+\frac{1}{6} \, (3 x_1 + 4 (x_2 + x_3 + x_4)) + 
 \frac{1}{6}  \,(-3 x_1^2 - 4 x_2^2- 4 x_3^2 + 10 x_3 x_4 \nonumber\\
 &- 4 x_4^2 + 
    10 x_2 (x_3 + x_4) + 5 x_1 (x_2 + x_3 + x_4))+\ldots\,,\\
    G_{10}(\bx)=\,&\frac{1}{3} \, (3 x_1 - x_2 - x_3 - x_4) + 
 \frac{1}{9}  \,(-9 x_1^2 - x_1 x_2 + 3 x_2^2 + 3 x_3^2 - 55 x_3 x_4\nonumber\\
 & + 
    3 x_4^2 + 41 x_1 (x_3 + x_4) - 13 x_2 (x_3 + x_4))+\ldots\,,\\
G_{11}(\bx)=\,&\frac{1}{3} \, (-x_1 + 3 x_2 - x_3 - x_4) + 
 \frac{1}{9}  \,(3 x_1^2 - 9 x_2^2 + 3 x_3^2 - 55 x_3 x_4 \nonumber\\
 &+ 3 x_4^2 + 
    41 x_2 (x_3 + x_4) - x_1 (x_2 + 13 (x_3 + x_4)))+\ldots\,,\\
G_{12}(\bx)=\,&\frac{1}{3} \, (-x_1 - x_2 + 3 x_3 - x_4) + 
 \frac{1}{9} \, (3 x_1^2 + 3 x_2^2 - 9 x_3^2\nonumber\\
 & + 
    x_1 (-13 x_2 + 41 x_3 - 55 x_4) + 41 x_3 x_4 + 3 x_4^2 - 
    x_2 (x_3 + 13 x_4))+\ldots\,,
        \end{align}
    \begin{align}
G_{13}(\bx)=\,&\frac{1}{3}  \,(-x_1 - x_2 - x_3 + 3 x_4) + 
 \frac{1}{9}  \,(3 x_1^2 + 3 x_2^2 - 55 x_2 x_3 \nonumber\\
 &+ 3 x_3^2 + 41 x_2 x_4 + 
    41 x_3 x_4 - 9 x_4^2 - x_1 (13 x_2 + 13 x_3 + x_4))+\ldots\,,
\end{align}
where the integration constants have been chosen to be zero.

\section{Boundary condition}
\label{app:init_cond}
A natural boundary condition for the differential equations are the values of the banana integrals in the small mass limit $\bx\to0$. In this section we discuss how we can obtain the leading term in the small-mass limit of the master integrals in the top sector (the subsectors are products of one-loop integrals). We focus on the master integrals $I_5,\ldots I_{15}$ in the initial basis. The initial condition for the canonical basis is easy to obtain.

We start by computing the boundary condition for the corner integral $I_{1,1,1,1,0,0,0,0,0}\,$. From~ref.~\cite{Bonisch:2021yfw}, we know that
\begin{align}
I_{1, \dots, 1,0,\dots,0}&(p^2, \bs{m}^2; 2 - 2\eps) =
-  \frac{e^{3\,\gamma_E\,\eps}}{\Gamma(1+3\eps)} e^{i\pi 3 \eps} 
\left( \frac{1}{p^2} \right)^{1 + 3\eps}\nonumber\\
&\sum_{j_k \in \{0,1\}} 
e^{i\pi (j - 1)\eps}
\frac{\Gamma(-\eps)^j \Gamma(\eps)^{4-j} \Gamma(1 + (j - 1)\eps)}{\Gamma(-j\eps)}\nonumber
\\
&\times 
\left[
\prod_{i=1}^{4} x_i^{(j_i - 1)\eps}
\sum_{n_k \in \mathbb{N}_0} 
\frac{(1 + j\eps)_{n} (1 + (j - 1)\eps)_{n}}{
\prod_{i=1}^{4} (1 + (-1)^{j_i + 1} \eps)_{n_i}}
\prod_{i=1}^{4} \frac{1}{n_i!} \left( x_i \right)^{n_i}
\right]\,,
\label{fullexp}
\end{align}
where $\bs{j}$ and $\bs{n}$ are four-component vectors, such that the entries $j_k$ of $\bs{j}$ are either 0 or 1 and $j=\sum_{k=1}^4 j_k$, while the entries $n_k$ of $\bs{n}$ take values in all of the natural numbers.
Expanding at the leading order in the kinematics (so taking $\bs{n}=\bs{0}$), we get 
\begin{align}
I_{1, \dots, 1,0,\dots,0}&(p^2, \bs{m}^2; 2 - 2\eps) =
-  \frac{e^{3\,\gamma_E\,\eps}}{\Gamma(1+3\eps)} e^{i\pi 3 \eps} 
\left( \frac{1}{p^2} \right)^{1 + 3\eps}
\times \nonumber\\
&\sum_{\bs{j} \in \{0,1\}} 
e^{i\pi (j - 1)\eps}
\frac{\Gamma(-\eps)^j \Gamma(\eps)^{4-j} \Gamma(1 + (j - 1)\eps)}{\Gamma(-j\eps)}
\prod_{i=1}^{4} x_i^{(j_i - 1)\eps}
+ \mathcal{O}(x_i^2)\,.
\end{align}

Let us now discuss the master integrals~$I_{2,1,1,1,0,0,0,0,0}\,,I_{1,2,1,1,0,0,0,0,0}\,,I_{1,1,2,1,0,0,0,0,0}\,,$ $I_{1,1,1,2,0,0,0,0,0}\,,I_{3,1,1,1,0,0,0,0,0}$. Since these integrals are derivatives of the corner integral (see eq.~\eqref{dercorner}), we can differentiate eq.~\eqref{fullexp} to get a boundary condition for them. 
Note that, since we are differentiating, to get the leading expansion in $x_i$ we cannot only take $\bs{n}=0$. For example, $I_{2,1,1,1,0,0,0,0,0}$ will receive, at leading order, contributions from $\bs{n}=(0,0,0,0)\,$ and $\bs{n}=(1,0,0,0)\,$.

Finally, let us discuss the master integrals $I_{10},\ldots, I_{14}$, which were defined using ISPs. 
We can build a different basis of master integrals using derivatives of $I_5$ only, so that the boundary condition can be computed using the results of ref.~\cite{Bonisch:2021yfw} for the small-mass limit. For example we may pick the integrals
\begin{equation}
    I_{2,2,1,1,0,0,0,0,0}\,,I_{2,1,2,1,0,0,0,0,0}\,,I_{2,1,1,2,0,0,0,0,0}\,,I_{1,2,2,1,0,0,0,0,0}\,,I_{1,1,2,2,0,0,0,0,0}\,.
\end{equation}
Their small-mass limit is easy to compute by taking derivatives of eq.~\eqref{fullexp}. Using IBP relations, we can relate those integrals to the master integrals defined in section~\ref{sec:setup}:
\begin{align}
&I_{1,1,1,1,-1,0,0,0,0}=\\
\nonumber&\frac{1}{2} (I_{0,1,1,1,0,0,0,0,0}+I_{1,1,0,1,0,0,0,0,0})+\frac{1}{6}( I_{1,0,1,1,0,0,0,0,0} +I_{1,1,1,0,0,0,0,0,0})\\
&+I_{1,1,1,1,0,0,0,0,0} \Big(\frac{-x_1+3 x_2-x_3+3 x_4-3}{12 \eps ^2}+\frac{-9 x_1+23 x_2-9 x_3+23 x_4-21}{12 \eps }\nonumber\\
&+\frac{1}{6} \left(-7 x_1+23 x_2-7 x_3+23 x_4-17\right)\Big)\nonumber+I_{1,1,1,2,0,0,0,0,0} \left(\frac{\left(x_4-1\right) x_4}{2 \eps ^2}+\frac{5 \left(x_4-1\right) x_4}{3 \eps }\right)\nonumber\\&+I_{1,1,2,1,0,0,0,0,0} \left(-\frac{x_3 \left(2 x_1-4 x_2+x_3-4 x_4+3\right)}{6 \eps ^2}-\frac{x_3 \left(5 x_1-7 x_2+2 x_3-7 x_4+4\right)}{3 \eps }\right)\nonumber\\&+\frac{2 x_3 \left(x_4-1\right) x_4 I_{1,1,2,2,0,0,0,0,0}}{3 \eps ^2}+I_{1,2,1,1,0,0,0,0,0} \left(\frac{\left(x_2-1\right) x_2}{2 \eps ^2}+\frac{5 \left(x_2-1\right) x_2}{3 \eps }\right)\nonumber\\
&+\frac{2 \left(x_2-1\right) x_2 x_3 I_{1,2,2,1,0,0,0,0,0}}{3 \eps ^2}+I_{2,1,1,1,0,0,0,0,0} \Big(\frac{1}{12 \eps ^2}(-3 x_1^2+14 x_2 x_1-10 x_3 x_1+14 x_4 x_1\nonumber\\
&-8 x_1+3 x_2^2-x_3^2+3 x_4^2-6 x_2+6 x_2 x_3-2 x_3-6 x_2 x_4+6 x_3 x_4-6 x_4+3)\nonumber\\
&+\frac{1}{12 \eps }(-9 x_1^2+34 x_2 x_1-26 x_3 x_1+34 x_4 x_1-18 x_1+3 x_2^2-x_3^2+3 x_4^2-6 x_2+6 x_2 x_3-2 x_3\nonumber\\
&-6 x_2 x_4+6 x_3 x_4-6 x_4+3)\Big)\nonumber\\
&+\frac{2 x_1 \left(x_4-1\right) x_4 I_{2,1,1,2,0,0,0,0,0}}{3 \eps ^2}+-\frac{x_1 x_3 \left(x_1-3 x_2+x_3-3 x_4+1\right) I_{2,1,2,1,0,0,0,0,0}}{3 \eps ^2}\nonumber\\
&+\frac{2 x_1 \left(x_2-1\right) x_2 I_{2,2,1,1,0,0,0,0,0}}{3 \eps ^2}-\frac{1}{6 \eps ^2}(x_1 \Big(x_1^2-6 x_2 x_1+6 x_3 x_1-6 x_4 x_1+2 x_1\nonumber\\
&-3 x_2^2+x_3^2-3 x_4^2+6 x_2-6 x_2 x_3+2 x_3+6 x_2 x_4-6 x_3 x_4+6 x_4-3\Big) I_{3,1,1,1,0,0,0,0,0})\,,\nonumber
\end{align}
\begin{align}
&I_{1,1,1,1,0,-1,0,0,0}=\\
&\frac{1}{6}(-I_{0,1,1,1,0,0,0,0,0}+ I_{1,0,1,1,0,0,0,0,0}-I_{1,1,0,1,0,0,0,0,0}+I_{1,1,1,0,0,0,0,0,0})\nonumber\\&+I_{1,1,1,1,0,0,0,0,0} \Big(\frac{-x_1-x_2-x_3-x_4+1}{12 \eps ^2}+\frac{-5 x_1-9 x_2-5 x_3-9 x_4+7}{12 \eps }\nonumber\\
&+\frac{1}{2} \left(-x_1-3 x_2-x_3-3 x_4+1\right)\Big)\nonumber\\&+I_{1,1,1,2,0,0,0,0,0} \left(-\frac{x_4 \left(x_1+x_3+x_4-1\right)}{6 \eps ^2}-\frac{x_4 \left(x_1+x_2+x_3+2 x_4-2\right)}{3 \eps }\right)\nonumber\\&+I_{1,1,2,1,0,0,0,0,0} \left(-\frac{x_3 \left(2 x_1+x_2+x_3+x_4-1\right)}{6 \eps ^2}-\frac{x_3 \left(2 x_1+2 x_2+x_3+2 x_4-1\right)}{3 \eps }\right)\nonumber\\
&-\frac{x_3 x_4 \left(3 x_1-x_2+x_3+x_4-1\right) I_{1,1,2,2,0,0,0,0,0}}{6 \eps ^2}\nonumber\\
&+I_{1,2,1,1,0,0,0,0,0} \left(-\frac{x_2 \left(x_1+x_2+x_3-1\right)}{6 \eps ^2}-\frac{x_2 \left(x_1+2 x_2+x_3+x_4-2\right)}{3 \eps }\right)\nonumber\\
&-\frac{x_2 x_3 \left(3 x_1+x_2+x_3-x_4-1\right) I_{1,2,2,1,0,0,0,0,0}}{6 \eps ^2}\nonumber
\end{align}
\begin{align}
&+I_{2,1,1,1,0,0,0,0,0} \Big(\frac{1}{12 \eps ^2}(-3 x_1^2-4 x_2 x_1-10 x_3 x_1-4 x_4 x_1+4 x_1-x_2^2-x_3^2-x_4^2\nonumber\\
\nonumber&+2 x_2-2 x_2 x_3+2 x_3+2 x_2 x_4-2 x_3 x_4+2 x_4-1\\
&+\frac{1}{12 \eps }(-5 x_1^2-10 x_2 x_1-14 x_3 x_1-10 x_4 x_1+6 x_1-x_2^2-x_3^2-x_4^2+2 x_2-2 x_2 x_3\nonumber\\
&+2 x_3+2 x_2 x_4-2 x_3 x_4+2 x_4-1)\Big)\nonumber\\
&-\frac{x_1 x_4 \left(x_1-x_2+3 x_3+x_4-1\right) I_{2,1,1,2,0,0,0,0,0}}{6 \eps ^2}-\frac{x_1 x_3 \left(x_1+x_2+x_3+x_4-1\right) I_{2,1,2,1,0,0,0,0,0}}{3 \eps ^2}\nonumber\\
&-\frac{x_1 x_2 \left(x_1+x_2+3 x_3-x_4-1\right) I_{2,2,1,1,0,0,0,0,0}}{6 \eps ^2}\nonumber\\
&-\frac{x_1}{6 \eps ^2} \Big(x_1^2+2 x_2 x_1+6 x_3 x_1+2 x_4 x_1-2 x_1+x_2^2+x_3^2+x_4^2\nonumber\\
&-2 x_2+2 x_2 x_3-2 x_3-2 x_2 x_4+2 x_3 x_4-2 x_4+1\Big) I_{3,1,1,1,0,0,0,0,0})\,,\nonumber\
\end{align}
\begin{align}
&I_{1,1,1,1,0,0,-1,0,0}=\\
&-\frac{1}{4} I_{0,1,1,1,0,0,0,0,0}+\frac{1}{12} (I_{1,0,1,1,0,0,0,0,0}+I_{1,1,0,1,0,0,0,0,0}+ I_{1,1,1,0,0,0,0,0,0})\nonumber\\&+I_{1,1,1,1,0,0,0,0,0} \Big(\frac{3 x_1-x_2-5 x_3-x_4+1}{24 \eps ^2}+\frac{27 x_1-9 x_2-37 x_3-9 x_4+7}{24 \eps }\nonumber\\
&+\frac{1}{4} \left(9 x_1-3 x_2-11 x_3-3 x_4+1\right)\Big)\nonumber\\
&+I_{1,1,1,2,0,0,0,0,0} \left(-\frac{x_4 \left(-2 x_1+4 x_3+x_4-1\right)}{12 \eps ^2}-\frac{x_4 \left(-5 x_1+x_2+7 x_3+2 x_4-2\right)}{6 \eps }\right)\nonumber\\&+I_{1,1,2,1,0,0,0,0,0} \left(\frac{\left(2 x_1-5 x_3+1\right) x_3}{12 \eps ^2}-\frac{x_3 \left(-5 x_1+x_2+8 x_3+x_4-2\right)}{6 \eps }\right)\nonumber\\
&+\frac{\left(x_2-x_3\right) x_3 x_4 I_{1,1,2,2,0,0,0,0,0}}{3 \eps ^2}+I_{1,2,1,1,0,0,0,0,0} \Big(-\frac{x_2 \left(-2 x_1+x_2+4 x_3-1\right)}{12 \eps ^2}\nonumber\\
&-\frac{x_2 \left(-5 x_1+2 x_2+7 x_3+x_4-2\right)}{6 \eps }\Big)\nonumber\\
&-\frac{x_2 x_3 \left(x_3-x_4\right) I_{1,2,2,1,0,0,0,0,0}}{3 \eps ^2}+I_{2,1,1,1,0,0,0,0,0} \Big(\frac{1}{24 \eps ^2}(9 x_1^2-6 x_2 x_1-18 x_3 x_1-6 x_4 x_1-x_2^2\nonumber\\
&-5 x_3^2-x_4^2+2 x_2-2 x_2 x_3+6 x_3+2 x_2 x_4-2 x_3 x_4+2 x_4-1)\nonumber\\
&+\frac{1}{24 \eps }(27 x_1^2-14 x_2 x_1-42 x_3 x_1-14 x_4 x_1-2 x_1-x_2^2-5 x_3^2-x_4^2+2 x_2-2 x_2 x_3\nonumber\\
&+6 x_3+2 x_2 x_4-2 x_3 x_4+2 x_4-1)\Big)\nonumber\\&+\frac{x_1 \left(x_1-x_2-3 x_3-x_4+1\right) x_4 I_{2,1,1,2,0,0,0,0,0}}{6 \eps ^2}+\frac{x_1 x_3 \left(x_1-x_2-3 x_3-x_4+1\right) I_{2,1,2,1,0,0,0,0,0}}{6 \eps ^2}\nonumber\\&+\frac{x_1 x_2 \left(x_1-x_2-3 x_3-x_4+1\right) I_{2,2,1,1,0,0,0,0,0}}{6 \eps ^2}+\frac{x_1}{12 \eps ^2} \Big(3 x_1^2-2 x_2 x_1-6 x_3 x_1-2 x_4 x_1-2 x_1-x_2^2\nonumber\\
&-5 x_3^2-x_4^2+2 x_2-2 x_2 x_3+6 x_3+2 x_2 x_4-2 x_3 x_4+2 x_4-1\Big) I_{3,1,1,1,0,0,0,0,0})\,,\nonumber
\end{align}
\begin{align}
&I_{1,1,1,1,0,0,0,-1,0}=\\
&\frac{1}{4} I_{1,0,1,1,0,0,0,0,0}-\frac{1}{12} (I_{0,1,1,1,0,0,0,0,0}+ I_{1,1,0,1,0,0,0,0,0}+ I_{1,1,1,0,0,0,0,0,0})\nonumber\\
&+I_{1,1,1,1,0,0,0,0,0} \Big(\frac{-x_1-x_2-x_3-x_4+1}{24 \eps ^2}+\frac{-5 x_1-13 x_2-5 x_3-5 x_4+7}{24 \eps }\nonumber\\
&+\frac{1}{4} \left(-x_1-5 x_2-x_3-x_4+3\right)\Big)\nonumber\\
&+I_{1,1,1,2,0,0,0,0,0} \left(-\frac{x_4 \left(2 x_1+2 x_3+x_4-1\right)}{12 \eps ^2}-\frac{x_4 \left(2 x_1+2 x_2+2 x_3+x_4-1\right)}{6 \eps }\right)\nonumber\\
&+I_{1,1,2,1,0,0,0,0,0} \left(-\frac{x_3 \left(2 x_1+x_3+2 x_4-1\right)}{12 \eps ^2}-\frac{x_3 \left(2 x_1+2 x_2+x_3+2 x_4-1\right)}{6 \eps }\right)\nonumber\\
&-\frac{x_3 x_4 \left(3 x_1-x_2+x_3+x_4-1\right) I_{1,1,2,2,0,0,0,0,0}}{6 \eps ^2}+I_{1,2,1,1,0,0,0,0,0} \left(-\frac{\left(x_2-1\right) x_2}{12 \eps ^2}-\frac{\left(x_2-1\right) x_2}{2 \eps }\right)\nonumber\\
&+I_{2,1,1,1,0,0,0,0,0} \Big(\frac{1}{24 \eps ^2}(-3 x_1^2+2 x_2 x_1-10 x_3 x_1-10 x_4 x_1+4 x_1-x_2^2-x_3^2\nonumber\\
&-x_4^2+2 x_2+2 x_2 x_3+2 x_3+2 x_2 x_4-6 x_3 x_4+2 x_4-1)\nonumber\\
&+\frac{1}{24 \eps }(-5 x_1^2-6 x_2 x_1-14 x_3 x_1-14 x_4 x_1+6 x_1\nonumber\\
&-x_2^2-x_3^2-x_4^2+2 x_2+2 x_2 x_3+2 x_3+2 x_2 x_4-6 x_3 x_4+2 x_4-1)\Big)\nonumber\\
&+-\frac{x_1 x_4 \left(x_1-x_2+3 x_3+x_4-1\right) I_{2,1,1,2,0,0,0,0,0}}{6 \eps ^2}+-\frac{x_1 x_3 \left(x_1-x_2+x_3+3 x_4-1\right) I_{2,1,2,1,0,0,0,0,0}}{6 \eps ^2}\nonumber\\&
-\frac{x_1 }{12 \eps ^2}(\Big(x_1^2-2 x_2 x_1+6 x_3 x_1+6 x_4 x_1-2 x_1+x_2^2+x_3^2+x_4^2\nonumber\\&
-2 x_2-2 x_2 x_3-2 x_3-2 x_2 x_4+6 x_3 x_4-2 x_4+1\Big) I_{3,1,1,1,0,0,0,0,0})\,,\nonumber
\end{align}
\begin{align}
&I_{1,1,1,1,0,0,0,0,-1}=\\
&\frac{1}{4} (I_{0,1,1,1,0,0,0,0,0}+ I_{1,0,1,1,0,0,0,0,0})-\frac{1}{12} (I_{1,1,0,1,0,0,0,0,0}+ I_{1,1,1,0,0,0,0,0,0})\nonumber\\
&+I_{1,1,1,1,0,0,0,0,0} \Big(\frac{x_1+x_2-3 x_3+x_4-1}{24 \eps ^2}+\frac{9 x_1+9 x_2-23 x_3+5 x_4-7}{24 \eps }\nonumber\\
&+\frac{1}{12} \left(13 x_1+13 x_2-23 x_3+x_4-7\right)\Big)+I_{1,1,1,2,0,0,0,0,0} \left(\frac{\left(x_4-1\right) x_4}{12 \eps ^2}+\frac{\left(x_4-1\right) x_4}{6 \eps }\right)\nonumber\\
&+I_{1,1,2,1,0,0,0,0,0} \left(-\frac{x_3 \left(3 x_3-4 x_4+1\right)}{12 \eps ^2}-\frac{x_3 \left(5 x_3-6 x_4+1\right)}{6 \eps }\right)\nonumber\\
&+\frac{x_3 \left(x_4-1\right) x_4 I_{1,1,2,2,0,0,0,0,0}}{3 \eps ^2}+I_{1,2,1,1,0,0,0,0,0} \left(\frac{x_2 \left(2 x_1+x_2-4 x_3-1\right)}{12 \eps ^2}+\frac{x_2 \left(5 x_1+2 x_2-7 x_3-x_4-2\right)}{6 \eps }\right)\nonumber\\
&-\frac{x_2 x_3 \left(x_3-x_4\right) I_{1,2,2,1,0,0,0,0,0}}{3 \eps ^2}+I_{2,1,1,1,0,0,0,0,0} \Big(\frac{1}{24 \eps ^2}(3 x_1^2+10 x_2 x_1-14 x_3 x_1-2 x_4 x_1-4 x_1+x_2^2\nonumber\\
&-3 x_3^2+x_4^2-2 x_2-6 x_2 x_3+2 x_3-2 x_2 x_4+10 x_3 x_4-2 x_4+1)+\frac{1}{24 \eps }(9 x_1^2+26 x_2 x_1-34 x_3 x_1\nonumber\\
&-6 x_4 x_1-10 x_1+x_2^2-3 x_3^2+x_4^2-2 x_2-6 x_2 x_3+2 x_3-2 x_2 x_4+10 x_3 x_4-2 x_4+1)\Big)\nonumber
\end{align}
\begin{align}
&+-\frac{x_1 x_3 \left(x_3-x_4\right) I_{2,1,2,1,0,0,0,0,0}}{3 \eps ^2}+\frac{x_1 x_2 \left(x_1+x_2-3 x_3-x_4-1\right) I_{2,2,1,1,0,0,0,0,0}}{6 \eps ^2}\nonumber\\&+\frac{x_1}{12 \eps ^2} \Big(x_1^2+6 x_2 x_1-6 x_3 x_1-2 x_4 x_1-2 x_1+x_2^2-3 x_3^2+x_4^2-2 x_2-6 x_2 x_3\nonumber\\
&+2 x_3-2 x_2 x_4+10 x_3 x_4-2 x_4+1\Big) I_{3,1,1,1,0,0,0,0,0})\,.\nonumber
\end{align}
\end{appendix}

\bibliographystyle{JHEP}
\bibliography{biblio.bib}

\end{document}